\documentclass[journal]{IEEEtran}
\usepackage{}

\usepackage{amsfonts}
\usepackage{amssymb}
\usepackage{amsmath}
\usepackage{mathrsfs}
\usepackage{amsmath,amssymb}
\usepackage{amsmath}
\usepackage{graphicx}
\usepackage{amsmath,amssymb,amsthm}
\usepackage{leftidx}
\usepackage{extarrows}
\usepackage{caption}
\usepackage{enumerate}
\usepackage{graphics}
\usepackage{subfigure}
\usepackage{float}
\usepackage{pict2e}
\usepackage{tikz}
\usepackage{cases}
\usepackage[numbers,sort&compress]{natbib}

\ifCLASSINFOpdf
\else
\fi
\begin{document}
%
\title{Stability Analysis for Stochastic Hybrid Inclusions: A Review}
%
%
%

\author{Hongye Su, Dandan Zhang
\thanks{This work was partially supported by National Key Research and Development Program of China (Grant No. 2018YFA0703803).
}
\thanks{H. Su and D. Zhang are with the
State Key Laboratory of Industrial Control Technology, the Institute of
Cyber-Systems and Control, the College of Control Science and Engineering, Zhejiang University, Hangzhou 310027,
China (e-mail: hysu@iipc.zju.edu.cn; zdandan3@zju.edu.cn).
}
}

\maketitle

\begin{abstract}

Stochastic hybrid inclusions (SHIs) address situations with the stochastic continuous evolution in a stochastic
differential inclusions and random jumps in the difference inclusions due to the forced (the state reaching
a boundary in the state space) and/or spontaneous (the state
vector may occur spontaneously) transitions. An obvious characteristic of SHIs is the non-uniqueness of random solutions, which can be ensured by the mild regularity conditions, as well as nominal robustness. Basic sufficient conditions for stability/recurrence in probability are usually expressed based on different types of Lyapunov
functions, including Lagrange/Lyapunov/Lyapunov-Forster functions respectively for Lagrange/Lyapunov/asymptotic stability in probability and Foster/Lagrange-Forster functions for recurrence, (weaker) relaxed Lyapunov-based sufficient conditions including Matrosov-Foster functions and the stochastic invariance principle, as well as Lyapunov-based necessary and sufficient conditions for asymptotic stability in probability or recurrence (i.e.,converse theorems), etc. The converse theorems involving smooth Lyapunov functions are guaranteed by the sequential compactness and thus robustness. In addition, the uniformity property and causality are analyzed for the stabilities in probability.
Hence, serving as a partial roadmap for the theoretical development of SHIs, also serving as inspiration, we anticipate that many of the open questions, including the prediction problem, the filtering problem and the control problem, will be resolved based on the techniques of SHIs.

\end{abstract}


%
\IEEEpeerreviewmaketitle


\section{Overview}



\IEEEPARstart{A}{fter} a series of mature theory results of deterministic hybrid systems \cite{2012teel}, which specify a flow/jump set where continuous evolution/instantaneous change may occur and a flow/jump map that determines how continuous evolution/instantaneous change may occur, the research on corresponding stochastic hybrid systems (SHSs) is almost fully mature due to their potential applications in a variety of practical problems in engineering \cite{3.1shs}. In SHS models, state changes are instantaneously triggered randomly in time and/or when states arrive a certain state space (i.e., forced transitions), where the jumps are determined via a probability distribution after instantaneous changes. A very general SHS model usually contains a diffusive component in the continuous evolutions \cite{5.2shs}, \cite{4.5shs}, or spontaneous transitions in the continuous evolutions \cite{3.1shs}, \cite{4.1shs}, \cite{7.3shs}.

This work surveys a very general SHS called stochastic hybrid inclusions (SHIs), where solutions are measurable mappings
from topological space $\Omega$ to the set of not identically empty outer semicontinuous mappings (set-valued mappings) from $\mathbb{R}^{2}$ to $\mathbb{R}^{n}$ \cite[pg. 182 \& Thm 5.50]{Rock1998}, where the values of the mapping are hybrid arcs with closed graphs as defined in \cite[Chap. 2]{2012teel} for example. These SHI models do not insist the boundedness of the expected value of the number of jumps in each
finite time interval, i.e., Zeno ensues as depicted in \cite[Fig. 2.3]{2012teel}, where hybrid arcs may be Zeno, eventually discrete, discrete, eventually continuous and continuous.

There exist two types of SHIs with forced transitions, according to whether there are spontaneous transitions \cite{3.1shs}, \cite{4.1shs}, \cite{7.3shs}, or not \cite{2.1shs}, \cite{3.1shs}, \cite{3.2shs}, \cite{3.3shs}, \cite{3.4shs}, \cite{3.5shs}, \cite{3.6shs}, \cite{3.7shs}, \cite{3.8shs}, \cite{4.2shs}, \cite{4.3shs}, \cite{4.4shs}, \cite{4.5shs}, \cite{4.6shs}, \cite{4.7shs}, \cite{5.1shs}, \cite{5.2shs}, \cite{5.4shs}, \cite{6.1shs}, \cite{7.1shs}, \cite{7.2shs}, \cite{7.3shs}. There exist three types of SHIs with forced transitions, according to set-valued mapping, i.e., only with the jump set-valued mapping (purely discrete-time systems) \cite{2.1shs}, \cite{3.2shs}, \cite{3.3shs}, \cite{3.5shs}, \cite{3.6shs}, \cite{3.8shs}, \cite{4.6shs}, \cite{4.7shs}, \cite{5.3shs}, \cite{5.4shs}, \cite{7.2shs}, simultaneously with flow/jump set-valued mappings \cite{3.1shs}, \cite{3.4shs}, \cite{3.7shs}, \cite{4.2shs}, \cite{4.3shs}, \cite{4.4shs}, \cite{5.1shs}, \cite{6.1shs}, \cite{7.1shs}, \cite{7.3shs}, with jump set-valued mappings and diffusion terms appearing in the flow set-valued mapping \cite{4.5shs, 5.2shs}. See Table \ref{2-3type}.

\begin{table*}
\centering  
\begin{tabular}{|l|c|c|}
\hline
~ & with spontaneous transitions & no spontaneous transitions \\ \hline
stochastic difference inclusions (SDIs):& ~ & \cite{2.1shs}, \cite{3.2shs}, \cite{3.3shs}, \cite{3.5shs}, \cite{3.6shs}\\
~~~~~~~~~~~~~~~~$x^{+}\in G(x, v)$  & ~ & \cite{3.8shs}, \cite{4.6shs}, \cite{4.7shs}, \cite{5.3shs}, \cite{5.4shs}, \cite{7.2shs}\\ \hline

stochastic differential and difference inclusions (SDDIs):& ~ & ~ \\
~~~~~~~~~~~~~~~~$\dot{x}\in F(x)$,~~~~~~~~~~~~~~~~~~ $x \in C$          &  \cite[Sec. 7]{3.1shs}, \cite{4.1shs}, \cite{7.3shs} & \cite{3.4shs}, \cite{3.7shs}, \cite{4.2shs}, \cite{4.3shs}, \cite{4.4shs}, \cite{5.1shs}, \cite{6.1shs}, \cite{7.1shs} \\
~~~~~~~~~~~~~~~~$x^{+}\in G(x, \nu^{+})$, ~~~~~~~~~~~$x \in D$  & ~&~\\ \hline

stochastic differential, diffusive and difference inclusions (SDDDIs):& ~ & ~ \\
~~~~~~~~~~~~~~~~$dx\in F(x)dt+B(x)dw$, $x\in C$  & ~ & \cite{4.5shs}, \cite{5.2shs} \\
~~~~~~~~~~~~~~~~$x^{+}\in G(x, v^{+})$, ~~~~~~~~~~~$x\in D$ & ~~&~ \\  \hline
\end{tabular}
 \caption{SHIs with forced transitions.}\label{2-3type}
\end{table*}

This survey serves as a partial roadmap for the theoretical development of SHIs, including the following properties:
1) A very general characteristic for SHIs, which may specify stochastic differential inclusions possibly with a diffusive component or spontaneous transitions, and difference inclusions.
2) Mild regularity conditions, in order to generate random solutions, robustness, the equivalence between non-uniform and uniform versions of stochastic stability properties, also to ensure the existence of a well-defined Lyapunov-based function for the stability in probability.
3) Non-unique random solutions, which are standard solutions as defined in \cite[Chap.2]{2012teel} with causal measurability properties that are adapted to the minimal filtration of stochastic variables. The non-uniqueness of random solutions comes from the potentially nontrivial overlap between the flow and jump sets, as well as the set-valued flow/jump mappings.
4) Stabilities in probability, such as Lyapunov-based sufficient conditions including Lagrange/Lyapunov/Lyapunov-Forster functions respectively for Lagrange/Lyapunov/asymptotic stability in probability and Foster/Lagrange-Forster functions for recurrence, (weaker) relaxed Lyapunov-based sufficient conditions including Matrosov-Foster functions and the stochastic invariance principle, as well as Lyapunov-based necessary and sufficient conditions for asymptotic stability in probability or recurrence (including converse theorems).
5) Sequential compactness for SHIs without diffusive component in the continuous evolutions, which is crucial to nominal robustness and thus converse Lyapunov theorems, also crucial to the connections between weakly totally recurrent in probability sets and invariant sets.
6) Nominal robustness of stability properties, i.e., stability in probability is
robust to sufficiently small, positive definite state perturbations, which is crucial for developing converse Lyapunov theorems. There exist three kinds of inflations for SHI models, i.e., slowing down the convergence, perturbations of system data, perturbations in the open bounded set (i.e., a smaller open set within the original set).
7) Causality, where non-causality means that inside information is leaked. The stabilities in probability may not hold if the causality constraint on solutions is relaxed; the nominal robustness may not hold if the perturbations, although arbitrarily small, are not strictly causal.
8) Converse theorems, which address Lyapunov-based necessary and sufficient conditions for asymptotic stability in probability or recurrence. Nominal robustness is used to smooth the established Lyapunov-based function in the final step.
Finally, we review a classical example on stochastic bouncing ball, in terms of the main theoretical results of stabilities in probability. Space constraints also limit our discussion on other SHS models' as shown in \cite[Sec. 2.2]{4.1shs}.

The paper is organized as follows: Section \ref{section2} describes a very general model for SHIs, characterizes random solutions under mild regularity conditions. Section \ref{section3} describes stability concepts and a series of Lyapunov-based conditions, including the stochastic invariance principle and Matrosov-Foster functions. Section \ref{section4} presents additional components of stability theory, as sequential compactness, nominal robustness, causality and converse theorems. Section \ref{section5} describes the applications and Section \ref{section6} concludes the paper.

\section{Preliminaries:}\label{section2}


\subsection{Basic notation and definitions}\label{app1}


We adopt the notation of \cite{3.1shs} and \cite{4.6shs}.
\textcolor[rgb]{0.00,0.00,0.00}{$\mathbb{R}_{\geq0}:=[0,\infty)$; $\mathbb{Z}_{\geq0}:= \{0, 1,\ldots\}$.} $\overline{\mathbb{R}_{\geq0}}=\mathbb{R}_{\geq0}\cup \{\infty\}$. $\overline{\hbox{con}}$ refers to the closed convex hull.
\textcolor[rgb]{0.00,0.00,0.00}{For a closed set $S\subset\mathbb{R}^{n}$ and $x\in\mathbb{R}^{n}$, $|x|_{S}:= \mbox{inf}_{y\in S} |x-y|$.} \textcolor[rgb]{0.00,0.00,0.00}{$\mathbb{B}$ (resp., $\mathbb{B}^{\circ}$) is the
closed (resp., open) unit ball in $\mathbb{R}^{n}$.}
\textcolor[rgb]{0.00,0.00,0.00}{For a closed set $S\subset \mathbb{R}^{n}$ and $\varepsilon > 0$, $S+\varepsilon\mathbb{B}$ (resp., $S+\varepsilon\mathbb{B}^{\circ}$) is the set $\{x\in \mathbb{R}^{n}: |x|_{S}\leq\varepsilon\}$ (resp., $\{x\in \mathbb{R}^{n}: |x|_{S}<\varepsilon\}$).}
\textcolor[rgb]{0.00,0.00,0.00}{The indicator function $\mathbb{I}_{S}: \mathbb{R}^{n}\rightarrow \{0, 1\}$ is defined by $\mathbb{I}_{S}(x)=1$ if and only if $x\in S$.}
\textcolor[rgb]{0.00,0.00,0.00}{A function $\phi: \mathbb{R}^{n}\rightarrow \mathbb{R}$ is \emph{upper semicontinuous} if $\limsup_{i\rightarrow\infty}\phi(x_{i})\leq\phi(x)$ whenever $\lim_{i\rightarrow\infty}x_{i}=x$.}
\textcolor[rgb]{0.00,0.00,0.00}{$\mathbb{I}_{S}$ is upper semicontinuous for closed $S$.}
\textcolor[rgb]{0.00,0.00,0.00}{Set-valued mappings (often just called ``mappings'') are indicated by the symbol $\rightrightarrows$.}
\textcolor[rgb]{0.00,0.00,0.00}{A mapping $M: \mathbb{R}^{p} \rightrightarrows \mathbb{R}^{n}$ is \emph{outer semicontinuous} if, for each $(x_{i}, y_{i})\rightarrow(x, y) \in \mathbb{R}^{p}\times\mathbb{R}^{n}$ satisfying $y_{i}\in M(x_{i})$ for all $i\in \mathbb{Z}_{\geq0}$, $y\in M(x)$.}
\textcolor[rgb]{0.00,0.00,0.00}{A mapping $M$ is \emph{locally bounded} if, for each bounded set $K\subset\mathbb{R}^{p}$, $M(K):=\bigcup_{x\in K}M(x)$ is bounded.}
\textcolor[rgb]{0.00,0.00,0.00}{$B(\mathbb{R}^{m})$ is the Borel $\sigma$-field on $\mathbb{R}^{m}$.}
\textcolor[rgb]{0.00,0.00,0.00}{Given a measure space $(\Omega,\mathcal {F})$, a set $F\subset \Omega$ is $\mathcal {F}$-\emph{measurable} if $F\in \mathcal {F}$.}
Let $(\Omega,\mathcal {F})$ be a measure space, e.g., $(\mathbb{R}^{m}, B(\mathbb{R}^{m}))$ where $B(\mathbb{R}^{m})$ is the
Borel field, i.e., the set of subsets of generated from open subsets of $\mathbb{R}^{m}$ through the algebra of complements and finite and countable unions.
\textcolor[rgb]{0.00,0.00,0.00}{A mapping $M: \mathbb{R}^{p} \rightrightarrows \mathbb{R}^{n}$ is $\mathcal {F}$-measurable \cite[Def. 14.1]{Rock1998} if, for each
open set $\mathcal {O}\subset\mathbb{R}^{n}$, $M^{-1}(\mathcal {O}):=\{v\in\mathbb{R}^{p}: M(v)\cap \mathcal {O}\neq\varnothing\}\in \mathcal {F}$.}
\textcolor[rgb]{0.00,0.00,0.00}{When the values of $M$ are closed, measurability is equivalent to $M^{-1}(\mathcal {C})$ being measurable for each closed set $C\subset \mathbb{R}^{n}$ \cite[Thm. 14.3]{Rock1998}.}
\textcolor[rgb]{0.00,0.00,0.00}{A function $\alpha: \mathbb{R}_{\geq0}\rightarrow\mathbb{R}_{\geq0}$ is \emph{of class} $\mathcal {K}$ if it is
continuous, strictly increasing and $\alpha(0)=0$.}
\textcolor[rgb]{0.00,0.00,0.00}{It is \emph{of class} $\mathcal {K}_{\infty}$ if it is of class $\mathcal {K}$ and unbounded.}
\textcolor[rgb]{0.00,0.00,0.00}{A function $\psi: \mathbb{Z}_{\geq0}\rightarrow\mathbb{R}_{\geq0}$ is \emph{of class} $\mathcal {L}$ if it is nonincreasing and $\lim_{\ell\rightarrow\infty}\psi(\ell)=0$.}
\textcolor[rgb]{0.00,0.00,0.00}{A function $\gamma: \mathbb{R}_{\geq0}\times\mathbb{Z}_{\geq0}\rightarrow\mathbb{R}_{\geq0}$ is \emph{of class} $\mathcal {K}\mathcal {L}$ if $\gamma(\cdot, k)\in\mathcal {K}$ for each $k\in \mathbb{Z}_{\geq0}$ and $\gamma(s,\cdot)\in \mathcal {L}$ for each $s\in \mathbb{R}_{\geq0}$.}
\textcolor[rgb]{0.00,0.00,0.00}{Given a compact set $\mathcal {A}\subset\mathbb{R}^{n}$, $\rho: \mathbb{R}^{n}\rightarrow\mathbb{R}_{\geq0}$ is \emph{of class} $\mathcal {P}\mathcal {D}(\mathcal {A})$ if it is continuous and $\rho(x)=0\Leftrightarrow x\in \mathcal {A}$.}
\textcolor[rgb]{0.00,0.00,0.00}{A function $\rho: \mathbb{R}_{\geq0}\rightarrow\mathbb{R}_{\geq0}$ is \emph{of class} $\mathcal {P}\mathcal {D}$ if it is continuous and $\rho(s)=0\Leftrightarrow s=0$.}

\subsection{Models}\label{section2.0}

We present a very general model for SHIs, usually containing diffusion terms and spontaneous transitions in the continuous evolutions by combining with \cite{4.5shs, 5.2shs} and \cite{4.1shs}, which can be specialized into the former two cases in Table \ref{2-3type}.

In order to cover all the types in Table \ref{2-3type}, a general model is written formally as
\begin{subequations}\label{4.5shs:1}
\begin{align}
&x\in C, ~~~dx\in F(x)dt+B(x)dw,~\uparrow\lambda(x)\label{4.5shs:1A}\\
&x\in D, ~~~x^{+}\in G(x, v^{+}),~~~~~~~~~~~~v\sim\mu(\cdot)\label{4.5shs:1B}
\end{align}
\end{subequations}
where $x\in \mathbb{R}^{n}$ is the state, $dx$ represents infinitesimal change
in the state (flowing), $x^{+}$ represents instantaneous change
in the state (jumping), $C$ is the flow set (where flowing
is allowed), $F$ is called the drift map, $B$ is called the
dispersion map, (the pair $(F,B)$ is called the flow map),
$D$ is the jump set (where jumps are allowed), and $G$ is the
jump map. The notation $v^{+}$ in (\ref{4.5shs:1B})
is a placeholder for a sequence of independent, identically
distributed (i.i.d.) input random variables $\textbf{v}_{i}: \Omega \rightarrow \mathbb{R}^{m}$, $i\in \{1,2,\cdots\}$, defined on a probability space $(\Omega,\mathcal {F},\mathbb{P})$. We use $\mu$ for the distribution function derived from these i.i.d.
random variables. The variable $w$ in (\ref{4.5shs:1A}) represents an $r$-dimensional Brownian motion (or Weiner process), defined on the same probability space. A measurable jump-rate function $\lambda :C\rightarrow \mathbb{R}_{\geq0}$ determines the probability of jumps at random times during flows, which is specialized to constant 1 in \cite{7.3shs}.

Often the data of the stochastic hybrid inclusion is
written as $\mathcal {H}:= (C, (F,B),D,G, \lambda, \mu)$, which can be specialized into $\mathcal {H}= (D, G, \lambda, \mu)$ for SDIs, $\mathcal {H}=(C, F,D,G, \lambda, \mu)$ for SDDIs and $\mathcal {H}= (C, (F,B),D,G, \lambda, \mu)$ for SDDDIs with ($\lambda\neq0$) and/or without ($\lambda=0$) spontaneous transitions, as shown in Table \ref{2-3type}. Note that \cite[Sec. 7]{3.1shs} and \cite{7.3shs} discuss the spontaneous transitions, while it is mentioned but not fully discussed in \cite{4.1shs}. So the main results throughout this paper are the case without spontaneous transitions, unless otherwise indicated.

\subsection{Stochastic hybrid basic conditions}\label{section2.1}

Likewise, we present the mild regularity conditions for SHS (\ref{4.5shs:1}), adopted from \cite{5.2shs} and \cite{4.5shs}.

\textbf{Assumption 1} (Stochastic hybrid basic conditions).
\begin{itemize}
  \item [A1)] The sets $C,D \subseteq \mathbb{R}^{n}$ are closed;
  \item [A2)] The mapping $F: \mathbb{R}^{n}\rightrightarrows\mathbb{R}^{n}$ is outer-semicontinuous, locally
bounded with nonempty convex values on $C$;
  \item [A3)] \begin{itemize}
             \item [3a)] The mapping $G: \mathbb{R}^{n}\times\mathbb{R}^{m}\rightrightarrows\mathbb{R}^{n}$ is locally bounded;
             \item [3b)] The mapping $v\mapsto \mbox{graph}(G(\cdot, v)):= \{(x, y)\in \mathbb{R}^{2n}: y\in G(x, v)\}$ is measurable (with respect to the Borel $\sigma$-algebra on $\mathbb{R}^{m}$) with closed values;
           \end{itemize}
  \item [A4)] $B:\mathbb{R}^{n}\rightrightarrows\mathbb{R}^{n\times r}$ is outer semicontinuous, locally
bounded and, for each $x\in C$, $B(x)$ is nonempty, convex and diagonally convex, i.e.,
the set $\{\sigma\in \mathbb{R}^{n}\times\mathbb{R}^{n}: \sigma=b^{T}b, b\in B(x)\}$ is convex.
\end{itemize}

\textbf{Remark 1.}
Now we explain the motivation of Assumption 1: The assumption of closed values in 3b) is equivalent to the outer semicontinuity of $v\mapsto G(x, v)$ for each $v\in \mathbb{R}^{n}$ according to \cite[Thm. 5.7(a)]{Rock1998}. These regularity conditions in Assumption 1 are used to: (i) generate random solutions to (\ref{4.5shs:1}) as established in \cite[Section IV]{4.5shs}, based on the conditions of local boundedness and measurability; (j) guarantee robustness when $B(x)=0$ and thus the equivalence between non-uniform and\footnotemark[1] uniform versions of stochastic stability properties, based on the conditions of local boundedness and outer semicontinuity used in for
example \cite[Claim 1, Thms. 2-3]{4.6shs} and \cite[Thms. 4-5]{3.2shs} for the difference inclusions, as well as \cite{7.1shs} for (\ref{4.5shs:1}). In particular, the existence of random solutions for SHS $\mathcal {H}$ (\ref{4.5shs:1}) can be ensured by the measurability in item 3b) and the locally bounded property in Assumption 1. The measurability in 3b) implies that $v\mapsto G(x, v)$ is measurable for any $x\in \mathbb{R}^{n}$ as analyzed in \cite[Prop. 2]{3.6shs} and \cite[Remark 1]{4.7shs} based on \cite[Thm. 14.13(b)]{Rock1998}, which can be also used to establish certain well-posed integrals used later.
Conversely, as analyzed in \cite{3.1shs} and \cite{3.3shs}, item 3b) holds if $G(x, v)$ is outer semicontinuous; holds if the domain
of $v\mapsto G(x, v)$ is countable; also holds if a single-valued mapping $G$ (i.e, a function) is a Carath\'{e}odory mapping \cite[Ex. 14.15]{Rock1998}, i.e., $x\mapsto G(x, v)$ is continuous and $x\mapsto G(x, v)$ is measurable.
Moreover, local boundedness and outer semicontinuity play an important role in the converse theorem such as \cite[Thm. 1]{4.6shs} and \cite[Thm. 1]{3.2shs} for difference inclusion, i.e., to ensure the existence of a continuous Lyapunov-Foster function for the stability in probability of SHS $\mathcal {H}$ (\ref{4.5shs:1}) as envisioned in \cite[Section 8]{7.1shs}, which, without those properties, may not exist even for non-stochastic systems \cite{4.6shs[12]}. Refer to \cite[Exam. 2]{4.2shs} that \cite[Thm. 6]{6.1shs} may fail if Assumption 1 does not hold.

\footnotetext [1]{Note that when spontaneous transitions exist (i.e., $\lambda\neq0$), Assumption 1 is not enough for ensuring the equivalence between non-uniform and uniform versions, see \cite[Sec. 4.6.2]{4.1shs}.}

A relationship route is presented in Fig. \ref{item3b} for item 3b) in Assumption 1. For other features of Assumption 1 which are not used explicitly here, for example, the diagonal convexity in item A4) will eventually play a role in existence results and/or sequential compactness results, which is featured extensively in \cite[Chapters 3-7, pg. 133-267]{MK2013}. Assumption 1 essentially agrees with the non-stochastic hybrid basic assumptions in \cite[Assumption 6.5]{2012teel}, if $B(x)=0$ and $G$ does not depend on $v$.

\begin{figure}[!htb]
\setlength{\unitlength}{0.7mm}
$$\begin{picture}(114,100)

\put(12,70){$v\mapsto \mbox{graph}(G(\cdot, v))$~~~~$v\mapsto \mbox{graph}(G(\cdot, v))$}
\put(13,64){~~~is measurable~~~~~~~~~is closed-valued}
\multiput(55,59)(0,1){20}{\line(0,1){0.5}}
\put(8,59){\line(1,0){94}}
\put(8,79){\line(1,0){94}}
\put(8,59){\line(0,1){20}}
\put(102,59){\line(0,1){20}}

\put(65,59){\vector(0,1){2}}
\put(65,59){\vector(0,-1){10}}
\put(67,51){\cite[Thm. 5.7 (a)]{Rock1998}}

\put(66,42){$x\mapsto G(x, v)$ is osc}
\put(62,49){\line(1,0){47}}
\put(62,39){\line(1,0){47}}
\put(62,39){\line(0,1){10}}
\put(109,39){\line(0,1){10}}

\put(106,49){\line(0,1){34}}
\put(106,83){\line(-1,0){101}}
\put(5,83){\vector(0,-1){33}}

\put(59,60){\line(1,0){2}}
\put(60,60){\vector(0,-1){30}}

\put(9,60){\line(1,0){2}}
\put(10,60){\vector(0,-1){10}}
\put(11,52){\cite[Thm. 14.13 (a)]{Rock1998}}

\put(8.5,43.5){$v\mapsto G(x, v)$}
\put(7.5,37.5){is measurable}
\put(1,35){\line(1,0){40}}
\put(1,50){\line(1,0){40}}
\put(1,35){\line(0,1){15}}
\put(41,35){\line(0,1){15}}

\put(31,35){\vector(0,-1){5}}
\put(3,23){$G(\cdot, v)$ is closed-valued and measurable}
\put(0,20){\line(1,0){90}}
\put(0,30){\line(1,0){90}}
\put(0,20){\line(0,1){10}}
\put(90,20){\line(0,1){10}}

\put(95,23){$V$ is usc}
\put(94,20){\line(1,0){20}}
\put(94,30){\line(1,0){20}}
\put(94,20){\line(0,1){10}}
\put(114,20){\line(0,1){10}}

\put(103,20){\vector(0,-1){10}}
\put(16,20){\vector(0,-1){10}}
\put(17,13){\cite[Exams. 14.30 \& 14.32, Thm. 14.37]{Rock1998}}
\put(11.45,3){$\int_{\mathbb{R}^{m}}\sup_{g\in G(x,v)}V(g)\mu(dv)$ is well-defined}
\put(-1,0){\line(1,0){116.5}}
\put(-1,10){\line(1,0){116.5}}
\put(-1,0){\line(0,1){10}}
\put(115,0){\line(0,1){10}}

\put(48,99){$G(x, v)$ is osc or}
\put(39.1,93){$\mbox{dom}(v\mapsto G(x, v))$ is countable}
\put(50,87){or $G(x, v)$ is a CM}
\put(38,105){\line(1,0){67}}
\put(38,85){\line(1,0){67}}
\put(38,85){\line(0,1){20}}
\put(105,85){\line(0,1){20}}

\put(3.5,89){\cite[Assum. 2]{3.1shs}}
\put(38,92.5){\line(-1,0){5}}
\put(33,92.6){\line(0,-1){8.24}}
\put(33,81.74){\vector(0,-1){5}}
\put(30.1,81.75){$\in$}
\end{picture}$$
\caption{A roadmap of the relationship about item 3b) in Assumption 1 \cite[Assumption 2 and Lemma 4.1]{3.1shs}, where $\int_{\mathbb{R}^{m}}\sup_{g\in G(x,v)}V(g)\mu(dv)$ will be used later for hybrid It\^{o}'s rule and hybrid Dynkin's formula, ``CM'' denotes Carath\'{e}odory mapping, ``osc'' denotes ``outer semicontinuous'' and ``usc'' denotes ``upper semicontinuous''.}\label{item3b}
\end{figure}


A function $V: \hbox{dom} V\rightarrow \mathbb{R}$ is a certification candidate for $\mathcal {H}= (C,(F,B),D,G, \mu)$, denoted as $V\in \mathcal {D}(\mathcal {H})$, i.e.,
\begin{itemize}
  \item [C1.] $C\cup D\cup G(D\times \mathcal {V})\subset \hbox{dom} V$;
  \item [C2.] $0\leq V(x)$ for all $x\in C\cup D\cup G(D\times \mathcal {V})$;
  \item [C3.] the quantity $\int_{\mathbb{R}^{m}}\sup_{g\in G(x,v)}V(g)\mu(dv)$ is well-defined for each
$x\in C\cup D$, using the convention that $\sup_{g\in G(x,v)}V(g) = 0$ when
$G(x,v) =\emptyset$, justified by the preceding item C2,
  \item [C4.] $V$ is twice continuously differentiable on an open set containing $C$,
\end{itemize}
where $\mathcal {V}:=\cup_{\omega\in \Omega, i\in \mathbb{Z}_{\geq1}}v_{i}(\omega)$, and $\{v_{i}\}_{i=1}^{\infty}$ are a sequence of i.i.d. random variables.

Given closed set $S\subset \mathbb{R}^{n}$, let $V\in \mathcal {D}(\mathcal {H}_{\cap S})$ be a candidate for $\mathcal {H}_{\cap S}:=(C\cap S,(F,B),D\cap S,G\cap S, \mu)$.

Based on Assumption 1 and \cite[Lem. 4.1]{3.1shs}, any upper semicontinuous function for $\mathcal {H}$ that satisfies C1-C2 is a certification candidate for $\mathcal {H}$. Tighter than upper semicontinuity, the certification candidate $V$ is a partially Lipschitz or partially $C^{1}$ for $\mathcal {H}= (C,F,D,G)$ if $V$ is locally Lipschitz (resp., smooth, or equivalently, continuously differentiable) on an open
set containing $C\setminus L_{V}(0)$ as well as continuous on $C$, where $L_{V}(0):=\{x\in\hbox{dom} V: V(x)=0\}$. Now we presented sufficient conditions for the uniform global asymptotic
stability in probability as shown in \cite[Thm. 4.5]{3.1shs}: the existence of a Lyapunov-Foster function implies uniform global asymptotic
stability in probability.

\subsection{Random solutions}\label{section2.1}

As presented in \cite{3.1shs},
Random solutions $x(\omega)$ to (\ref{4.5shs:1}) have two features: 1) For each $\omega\in \Omega$, the sample path $x(\omega)$, appropriately coupled with the random input sequence $v(\omega)$, must be a standard solution; we adopt the solution concept of \cite[Chap.2]{2012teel}. 2) $\omega \mapsto x(\omega)$ has measurability properties that are adapted to the minimal filtration of $v$. This measurability is used to express stability and recurrence in terms of probabilities and to enforce causal dependence on $v$.
Measurability concepts are from set-valued analysis \cite[Chap.14]{Rock1998}, causality is related to non-anticipative control laws in stochastic control, i.e., no inside information, see \cite[Sec. 6.2]{3.1shs}.
Regardless of which form the model (\ref{4.5shs:1}) degenerates into in the Table \ref{2-3type}, both of the two features must be followed, which have been summarized into pathwise feasibility and causal measurability, see \cite{4.6shs, 3.3shs, 4.7shs} or \cite{3.2shs, 3.5shs} for more details on SDIs, \cite{6.1shs} for SDDIs, \cite{7.3shs} for SDDIs with spontaneous transitions, and \cite{4.5shs} for SDDDIs.

The potentially nontrivial overlap between the flow and jump sets, as well as the set-valued flow/jump mappings imply the non-uniqueness of solutions in these SHI models. Here we borrow the analysis in \cite{4.5shs} for SDDDIs without spontaneous transitions (i.e., $\lambda=0$).


\indent 1) \emph{Random inputs, Brownian motion, hybrid filtrations and hybrid stopping times:} On a given probability space $(\Omega,\mathcal {F}, \mathbb{P})$, define a sequence of i.i.d. random variables $\{\textbf{v}_{i}\}_{i=1}^{\infty}:=\textbf{v}$ with $\textbf{v}_{i}: \Omega \rightarrow \mathbb{R}^{m}$, which drives the jumps of the
hybrid system (\ref{4.5shs:1}). Define $\hat{\mathcal {F}}_{0}:=\{\Omega, \varnothing\}$. Denote $\{\hat{\mathcal {F}}_{i}\}_{i=1}^{\infty}$ as the minimal filtration associated to the random process $\{\textbf{v}_{i}\}_{i=1}^{\infty}$: $\hat{\mathcal {F}}_{i}$ is a sub-$\sigma$-field of $\mathcal {F}$, generated by $(\textbf{v}_{1}(\omega),\cdots,\textbf{v}_{i}(\omega))$ \cite[Def. 4, pg. 171]{Fristedt1997}. Define $\hat{\mathbb{F}}:=\{\hat{\mathcal {F}}_{i}\}_{i\in \mathbb{Z}_{\geq0}}$. Independent of $\hat{\mathbb{F}}$, the additional filtration $\tilde{\mathbb{F}}:=\{\tilde{\mathcal {F}}_{t}\}_{t\in \mathbb{R}_{\geq0}}$ is such that $\tilde{\mathcal {F}}_{0}$ contains all $\mathbb{P}$-negligible sets in $\mathcal {F}$. Note that this requirement does not means that $\tilde{\mathcal {F}}_{0}$ is complete \cite[pg. 1]{MK2013}, \cite{4.2shs[4]}, since some $\mathbb{P}$-negligible sets in $\mathcal {F}$ may not be in the completion of $\tilde{\mathcal {F}}_{0}$ \cite[pg. 4]{IK1988}.

Given the filtered probability space $(\Omega,\mathcal {F}, \tilde{\mathbb{F}}, \mathbb{P})$, an $\tilde{\mathbb{F}}$-adapted
Brownian motion relative to this space is a continuous process $\textbf{w}=\{\textbf{w}(t,\cdot)\}_{t\geq0}$ with the properties that
$\sigma(\textbf{w}(s,\cdot), s\in[0, t])\subset \tilde{\mathcal {F}}_{t}$ for all $t\geq0$, $\textbf{w}(0,\cdot)=0$ almost
surely and for $0\leq s<t$ the increment $\textbf{w}(t,\cdot)-\textbf{w}(s,\cdot)$ is normally distributed with mean zero and variance $t-s$ and is independent of $\tilde{\mathcal {F}}_{s}$ \cite[Def. 1.1, pg. 47]{IK1988}.

Simultaneously equipped with the two independent filtration, i.e., the discrete-time case $\hat{\mathbb{F}}$ and the
continuous-time case $\tilde{\mathbb{F}}$, the hybrid filtration can be defined as
\begin{equation}\label{4.5shs:1.0}
\begin{split}
\mathbb{F}:=\left\{\mathcal {F}_{t,j}:= \bigcap_{\varepsilon>0}\sigma(\tilde{\mathcal {F}}_{t+\varepsilon}, \hat{\mathcal {F}}_{j})\right\}_{(t,j)\in \mathbb{R}_{\geq0}\times \mathbb{Z}_{\geq0}}.
\end{split}
\end{equation}
For the hybrid filtration, let $(s, i)\preceq(t, j)$ denote $s\leq t$ and $i\leq j$, respectively with $\tilde{\mathcal {F}}_{s}\subset\tilde{\mathcal {F}}_{t}$ \cite[pg. 7]{KLC1990} and $\hat{\mathcal {F}}_{i}\subset\hat{\mathcal {F}}_{j}$
\cite[Def. 4, pg. 171]{Fristedt1997}, thus resulting in $\mathcal {F}_{s,i}\subset\mathcal {F}_{t,j}$ as defined in \cite{4.5shs}. As stated in \cite[Lemma 1]{5.2shs}, for the hybrid filtration $\mathbb{F}$ in (\ref{4.5shs:1.0}), $\mathcal {F}_{0,0}$ contains all $\mathbb{P}$-negligible events in $\mathcal {F}$ and is right-continuous in $t$ \cite{4.5shs}, i.e., $\mathcal {F}_{t,j} = \cap_{\varepsilon>0}\mathcal {F}_{t+\varepsilon,j}$ for all $(t,j)\in\mathbb{R}_{\geq0}\times \mathbb{Z}_{\geq0}$. For each $\tilde{\mathbb{F}}$-adapted Brownian motion, the increment $\textbf{w}(t,\cdot)-\textbf{w}(s,\cdot)$ ($0\leq s< t$) is independent of $\tilde{\mathcal {F}}_{s,t}$ for each $j\in \mathbb{Z}_{\geq0}$. A \emph{hybrid stopping time
relative to} $\mathbb{F}$ is a pair of $\mathcal {F}$-measurable mappings $\textbf{T}: \Omega\rightarrow (\mathbb{R}_{\geq0}\cup\{\infty\})$ and $\textbf{J}: \Omega\rightarrow (\mathbb{Z}_{\geq0}\cup\{\infty\})$, if
\begin{equation*}\label{4.5shs:1.01}
\begin{split}
\{(\textbf{T}, \textbf{J})\preceq(t,j)\}\in \mathcal {F}_{t,j}, ~\forall(t,j)\in \mathbb{R}_{\geq0}\times \mathbb{Z}_{\geq0}.
\end{split}
\end{equation*}
As stated in \cite[Lemma 2]{5.2shs}, given two hybrid stopping times $(\textbf{T}_{1}, \textbf{J}_{1})$ and
$(\textbf{T}_{2}, \textbf{J}_{2})$, then $(\textbf{T}_{1}\vee\textbf{T}_{2}, \textbf{J}_{1}\vee\textbf{J}_{2})$ and $(\textbf{T}_{1}\wedge\textbf{T}_{2}, \textbf{J}_{1}\wedge\textbf{J}_{2})$ are also hybrid
stopping times, if,
\begin{equation}\label{4.5shs:1.02}
\begin{split}
(\textbf{T}_{1}, \textbf{J}_{1})\preceq
(\textbf{T}_{2}, \textbf{J}_{2})~\mbox{or}~
(\textbf{T}_{2}, \textbf{J}_{2})\preceq
(\textbf{T}_{1}, \textbf{J}_{1})
\end{split}
\end{equation}
for almost every $\omega\in \Omega$.

For a given hybrid filtration $\mathbb{F}$, a mapping $\textbf{H}$ defined on $\Omega$ is an $\mathbb{F}$-\emph{adapted
stochastic hybrid time domain}, if $\textbf{H}(\omega)$ is a hybrid time domain \cite[Chapter 2]{2012teel} for each $\omega\in \Omega$.
Thus, actually, the above condition (\ref{4.5shs:1.02}) always holds if $(\textbf{T}_{1}, \textbf{J}_{1})$ and $(\textbf{T}_{2}, \textbf{J}_{2})$ belong to the same hybrid time domain almost surely. The set-valued mapping $\omega\mapsto \overline{\textbf{H}(\omega)}\cap([0, t]\times\{0,\ldots, j\})$ is $\mathcal {F}_{t,j}$-measurable for each $(t,j)\in \mathbb{R}_{\geq0}\times \mathbb{Z}_{\geq0}$, and
\begin{equation}\label{4.5shs:1.03}
\begin{split}
(\textbf{T}_{\dashv}(\omega),\textbf{J}_{\dashv}(\omega)) :=\sup\{(t, j)\in\textbf{H}(\omega)\}
\end{split}
\end{equation}
is a hybrid stopping time.
Define the mappings $\textbf{T}_{i}: \Omega\rightarrow\mathbb{R}_{\geq0}$ as
\begin{equation}\label{4.5shs:1.04}
\begin{split}
\textbf{T}_{i}(\omega):=\inf\{t\in \mathbb{R}_{\geq0}: (t, i)\in\overline{\textbf{H}(\omega)}\},~\forall~i\in \mathbb{Z}_{\geq0},
\end{split}
\end{equation}
where $\textbf{H}$ is an $\mathbb{F}$-adapted stochastic hybrid time domain. As stated in \cite[Lemma 3]{5.2shs}, for each $i\in \mathbb{Z}_{\geq0}$, both $(\textbf{T}_{i}, i)$ and $(\textbf{T}_{i+1}, i)$ are hybrid stopping times, particularly $\{s\in [\textbf{T}_{i},\textbf{T}_{i+1})\}\in \mathcal {F}_{s,i}$. Thus, for the hybrid
stopping times $\{(\textbf{T}_{i}, i)\}_{i\in\mathbb{Z}_{\geq0}}$ resulting from an $\mathbb{F}$-adapted
stochastic hybrid time domain $\textbf{H}$, the $\sigma$-algebra $\mathcal {F}_{\textbf{T}_{i}, i}$ of events determined prior to the
stopping time $(\textbf{T}_{i}, i)$ is defined to consist of events $A\in\mathcal {F}$ in \cite{5.2shs} such that
\begin{equation*}\label{4.5shs:1.05}
\begin{split}
A\cap \{(\textbf{T}_{i+1}, i)\preceq(t, j)\}\in \mathcal {F}_{t,j},
\end{split}
\end{equation*}
for each $(t,j)\in \mathbb{R}_{\geq0}\times \mathbb{Z}_{\geq0}$. As stated in \cite[Lemma 4]{5.2shs}, for each $i\in\mathbb{Z}_{\geq0}$, $\mathcal {F}_{\textbf{T}_{i+1}, i}$ is a $\sigma$-algebra, $\textbf{T}_{i+1}$ is $\mathcal {F}_{\textbf{T}_{i+1}, i}$-measurable, and $\{\mathcal {F}_{i}\}_{i=0}^{\infty}$ with $\mathcal {F}_{i}:=\mathcal {F}_{\textbf{T}_{i+1}, i}$ is a filtration of the probability space $(\Omega,\mathcal {F}, \mathbb{P})$.

\indent 2) \emph{Solutions:} The following presented solutions' concepts to (\ref{4.5shs:1}) are all borrowed from \cite{5.2shs} and \cite{4.5shs}, which are evolved from the non-stochastic concepts for deterministic hybrid systems in \cite[Chapter 2]{2012teel}. Different from the non-stochastic case where a hybrid arc satisfying certain conditions is a solution candidate, here we mainly focus on a relaxed hybrid arc for a sample path of a solution candidate in the stochastic case.

\emph{A relaxed hybrid arc} is a function $\phi: \mbox{dom}(\phi)\rightarrow \mathbb{R}^{n}$ such that:
\begin{itemize}
  \item [1)] $\mbox{dom}(\phi)$ is a hybrid time domain (as defined in \cite[Def. 3, pg. 27]{2012teel});
  \item [2)] regarded as a set-valued mapping from $\mathbb{R}^{2}$ to $\mathbb{R}^{n}$, $\phi$ is outer semicontinuous (as defined in \cite[Def. 5.9, pg. 102]{2012teel}), i.e., for all $i\in \mathbb{Z}_{\geq0}$ and for each sequence $(t_{i}, j_{i}, z_{i})\rightarrow (t,j,z)$ satisfying $z_{i}=\phi(t_{i}, j_{i})$, we have $z=\phi(t, j)$.
\end{itemize}
In fact, \cite[Thm. 5.7(a), pg. 154]{Fristedt1997} implies that the set $\mbox{graph}(\phi)$ is closed where $\mbox{graph}(\phi):=\{(t,j,z)\in \mathbb{R}^{n+2}: z=\phi(t, j)\}$. For each $i\in \mathbb{Z}_{\geq0}$, item 2) also implies that the mapping
$\phi(\cdot, i): \mathcal {I}_{i}\rightarrow \mathbb{R}^{n}$ is continuous if $\mbox{dom}(\phi)\cap(\mathbb{R}_{\geq0}\times \{i\})=: \mathcal {I}_{i}\times \{i\}$ is nonempty. In contrast, the modifier ``relaxed'' here is used to emphasize the difference from the fact that, a hybrid arc for deterministic hybrid systems is such that for each $i\in \mathbb{Z}_{\geq0}$, $t\mapsto \phi(t,i)$ is locally absolutely continuous \cite[Def. 2.4, pg. 28]{2012teel}. Thus, we can correspondingly define \emph{a stochastic hybrid arc}, i.e., a mapping $\textbf{x}$ defined on $\Omega$ satisfying
\begin{itemize}
  \item for each $\omega\in \Omega$, $\textbf{x}(\omega)$ (abbr. $\textbf{x}_{\omega}$) is a relaxed hybrid arc;
  \item the set-valued mapping $\textbf{x}: \Omega\rightarrow \mathbb{R}^{n+2}$, i.e.,
\begin{equation*}\label{4.5shs:1.1}
\begin{split}
\omega&\mapsto \mbox{graph}(\textbf{x}(\omega)):=\\
&\{(t,j,z)\in\mathbb{R}^{n+2}: (t,j)\in \mbox{dom}~\textbf{x}_{\omega}, z=\textbf{x}_{\omega}(t,j)\}
\end{split}
\end{equation*}
is $\mathcal {F}$-measurable with closed values.
\end{itemize}
Correspondingly, an $\mathcal {F}$-\emph{adapted stochastic hybrid arc} is a stochastic hybrid arc $\textbf{x}$ where the set-valued mapping
\begin{equation*}\label{4.5shs:1.2}
\begin{split}
\omega\mapsto \mbox{graph}(\textbf{x}(\omega))\cap([0,t]\times&\{0,1,\ldots,j\}\times \mathbb{R}^{n})\\
:=&\mbox{graph}_{\preceq(t,j)}(\textbf{x}(\omega))
\end{split}
\end{equation*}
is $\mathcal {F}_{t,j}$-measurable for each $(t,j)\in \mathbb{R}_{\geq0}\times \mathbb{Z}_{\geq0}$.

\emph{A partial hybrid signal generator} is a mapping $\tilde{\phi}: \mbox{dom}(\tilde{\phi})\rightarrow \mathbb{R}^{n\times r}$ such that $\mbox{dom}(\tilde{\phi})$ is a subset of a hybrid time domain \cite[Def. 3, pg. 27]{2012teel}. Thus one can correspondingly define \emph{a stochastic partial hybrid signal generator} for a given measure space $(\Omega,\mathcal {F})$, i.e., a mapping $\textbf{f}$ satisfying
\begin{itemize}
 \item for each $\omega\in \Omega$, $\textbf{f}(\omega)$ (abbr. $\textbf{f}_{\omega}$) is a partial hybrid signal generator;
  \item viewed as a set-valued mapping, $\textbf{f}: \Omega\times \mathbb{R}_{\geq0}\times \mathbb{Z}_{\geq0}\rightarrow \mathbb{R}^{n\times r}$ is measurable.
\end{itemize}
Then, for a filtered probability space $(\Omega,\mathcal {F}, \mathbb{F}, \mathbb{P})$, an $\mathbb{F}$-\emph{adapted stochastic hybrid signal generator} is a stochastic partial hybrid signal generator satisfying for each $(s,i)\in \mathbb{R}_{\geq0}\times \mathbb{Z}_{\geq0}$, $\omega\mapsto \textbf{f}_{\omega}(s,i)$ is $\mathcal {F}_{s,i}$-measurable.

Following the above presented concepts, now we present the solution concept for SHS (\ref{4.5shs:1}), with a sequence of i.i.d. random variables $\textbf{v}:=\{\textbf{v}_{i}\}_{i=1}^{\infty}$. That is, a triple $(\textbf{x}, (\textbf{f},\textbf{b},\textbf{g}), \textbf{w})$ is a solution to the SHS (\ref{4.5shs:1}) starting at $x\in \mathbb{R}^{n}$ such that:
\begin{itemize}
  \item $\textbf{x}$ is an $\mathbb{F}$-adapted stochastic hybrid arc,
  \item $(\textbf{f},\textbf{b},\textbf{g})$ are $\mathbb{F}$-adapted stochastic partial hybrid signal
generators,
  \item $\textbf{w}$ is an $\tilde{\mathbb{F}}$-adapted Brownian motion,
\end{itemize}
such that $\textbf{x}(\omega)$ is a solution to (\ref{4.5shs:1}) for almost every $\omega\in \Omega$ selected by $(\textbf{f} (\omega), \textbf{b}(\omega), \textbf{g}(\omega))$, starting at $x$ with inputs $\textbf{w}(\cdot,\omega)$ and $\textbf{v}(\omega)$. Specifically, for almost every $\omega\in \Omega$,
\begin{itemize}
  \item [1)] $\textbf{x}_{\omega}(0, 0)=x$;

  \item [2)] for each $j\in\mathbb{Z}_{\geq0}$, if $I^{j}(\omega):= \{t : (t, j)\in \mbox{dom}~\textbf{x}(\omega)\}$ has nonempty interiors then, for every $t\in I^{j}(\omega)$,
\begin{itemize}
  \item [a)] $\textbf{x}_{\omega}(t, j)\in C$,
  \item [b)] \begin{itemize}
               \item [i)] $(t, j)\in \mbox{dom}~ \textbf{f}(\omega)\cap  \mbox{dom}~ \textbf{b}(\omega)$,
               \item [ii)] $\textbf{f}_{\omega}(t, j)\in F(\textbf{x}_{\omega}(t, j))$,
$\textbf{b}_{\omega}(t, j)\in B(\textbf{x}_{\omega}(t, j))$,
               \item [iii)] $\textbf{x}_{\omega}(t, j)-\textbf{x}_{\omega}(t_{j}, j)\\=\int^{t}_{T_{j}}\textbf{f}_{\omega}(s, j)ds+\int^{t}_{T_{j}}\textbf{b}_{\omega}(s, j)d \textbf{w}(s,\omega)$, where $\textbf{T}_{j}$ is defined in (\ref{4.5shs:1.04}) with $\textbf{H} = \mbox{dom}~\textbf{x}$;
             \end{itemize}
\end{itemize}

  \item [3)] if $(t,j), (t, j+1)\in\mbox{dom}~ \textbf{x}(\omega)$ then
\begin{itemize}
  \item [a)] $\textbf{x}_{\omega}(t, j)\in D$,
  \item [b)] $(t, j)\in \mbox{dom}~ \textbf{g}(\omega)$,
  \item [c)] $\textbf{g}_{\omega}(t, j)\in G(\textbf{x}_{\omega}(t, j), \textbf{v}_{j+1}(\omega))$,
  \item [d)] $\textbf{x}_{\omega}(t, j+1)=\textbf{g}_{\omega}(t, j)$.
\end{itemize}
\end{itemize}
Denote $\textbf{x}\in \mathcal {S}_{r}(K)$ as a solution $(\textbf{x}, (\textbf{f},\textbf{b},\textbf{g}), \textbf{w})$ starting at certain point in $K$. Thus, as stated in \cite[Lemma 5]{5.2shs}, we can get the following hybrid stopping times for a given solution $\textbf{x}$, i.e.,
\begin{itemize}
  \item the first time exiting from a closed set $\mathcal {C}$:
\begin{equation}\label{4.5shs:1.3}
\begin{split}
(\textbf{T}_{\mathcal {C}}(\omega), \textbf{J}_{\mathcal {C}}(\omega)) :=\inf \{(t, j)\in  \mbox{dom}~\textbf{x}(\omega): \textbf{x}_{\omega}(t, j)\notin\mathcal {C}\},
\end{split}
\end{equation}
  \item and the smallest time greater than $\tau\geq0$:
\begin{equation}\label{4.5shs:1.4}
\begin{split}
(\textbf{T}_{\tau}(\omega), \textbf{J}_{\tau}(\omega)) :=\inf \{(t, j)\in  \mbox{dom}~\textbf{x}(\omega): t+j>\tau\},
\end{split}
\end{equation}
\end{itemize}
As stated in \cite[Lemma 6]{5.2shs}, given solution $\textbf{x}$ with $\mbox{dom}~\textbf{x}:=\textbf{H}$,
viewed as a set-valued mapping with the empty value for each $\omega\in \Omega$ and each $i\in \mathbb{Z}_{\geq0}$ such that $\textbf{T}_{i+1}(\omega)=\infty$, both the (set-valued) mapping $\omega\mapsto
\mbox{graph}_{\preceq(\infty,i)}(\textbf{x}(\omega))$ and the mapping $\omega\mapsto \textbf{x}_{\omega}(\textbf{T}_{i+1}(\omega), i)$ are $\mathcal {F}_{i}$-measurable. Moreover, $\{\textbf{J}\leq i\}\in\mathcal {F}_{i}$ if $(\textbf{T}, \textbf{J})$ is an almost surely finite stopping time with $(\textbf{T}(\omega), \textbf{J}(\omega))\in\mbox{dom}~\textbf{x}(\omega)$ for almost all $\omega\in \Omega$.

\indent 3) \emph{Hybrid It\^{o}'s Rule and Dynkin's Formula:}
Given a solution $\textbf{x}$ with its hybrid stopping time $(\textbf{T}, \textbf{J})$ (i.e., $(\textbf{T}(\omega), \textbf{J}(\omega))\in\mbox{dom}~\textbf{x}(\omega)$ for almost all $\omega\in \Omega$). Define $\textbf{J}_{s}(\omega) := \sup \{j\in \mathbb{Z}_{\geq0}: (s, j)\in  \mbox{dom}~\textbf{x}\}$ for each $s\in[0,\textbf{T}(\omega)]$.

Given closed set $S\subset\mathbb{R}^{n}$. Define $\mathcal {H}_{\cap S} := (C\cap S, (F,B),D\cap S,G\cap S)$, thus $\mathcal {H} := \mathcal {H}_{\cap\mathbb{R}^{n}}$. Let $V\in \mathcal {D}(\mathcal {H}_{\cap S})$ when $V: \mathbb{R}^{n}\rightarrow \mathbb{R}$ is continuous on closed set $D\cap S$ and twice continuously differentiable on closed set $C \cap S$. Now we present hybrid It\^{o}'s rule and Dynkin's formula borrowed from \cite[Lemmas 7-9]{4.5shs}.

\indent \emph{Hybrid It\^{o}'s Rule} \cite[Lemma 7]{4.5shs}: Let $S\subset\mathbb{R}^{n}$ be closed, let
$V\in \mathcal {D}(\mathcal {H}_{\cap S})$, let $(\textbf{x}, (\textbf{f},\textbf{b},\textbf{g}), \textbf{w})$ be a solution of $\mathcal {H}_{\cap S}$ starting at $x$ and let $(\textbf{T}, \textbf{J})$ be a hybrid stopping time of $\textbf{x}$.
Then, almost surely,
\begin{equation*}\label{4.5shs:1.5}
\begin{split}
&V(\textbf{x}(\textbf{T}, \textbf{J}))-V(x)\\
=&\sum_{i=0}^{\textbf{J}-1}(V(\textbf{g}(\textbf{T}_{i+1},i))-V(\textbf{x}(\textbf{T}_{i+1},i)))\\
&+\int_{0}^{\textbf{T}}\langle \nabla V(\textbf{x}(s,\textbf{J}_{s})), \textbf{f}(s,\textbf{J}_{s})ds+\textbf{b}(s,\textbf{J}_{s})d\textbf{w}(s)\rangle\\
&+\int_{0}^{\textbf{T}}\frac{1}{2}\mbox{trace}(\textbf{b}(s,\textbf{J}_{s})\textbf{b}^{T}(s,\textbf{J}_{s})\nabla^{2} V(\textbf{x}(s,\textbf{J}_{s})))ds.
\end{split}
\end{equation*}

\indent \emph{Hybrid Dynkin's formula, part I-II} \cite[Lemmas 8-9]{4.5shs}:
Let Assumption 1 hold, let $S\subset\mathbb{R}^{n}$ be closed, let
$V\in \mathcal {D}(\mathcal {H}_{\cap S})$, let $(\textbf{x}, (\textbf{f},\textbf{b},\textbf{g}), \textbf{w})$ be a solution of $\mathcal {H}_{\cap S}$ starting at $x$ and let $(\textbf{T}, \textbf{J})$ be a hybrid stopping time of $\textbf{x}$. Suppose $\mathbb{E}[\textbf{T}] < \infty$ and that either $V$ has compact support or $(\textbf{T}, \textbf{J})$ is bounded by the first exit time from a compact set. Also suppose that $\textbf{J}$ is uniformly bounded almost surely. Then,
\begin{equation}\label{4.5shs:1.6}
\begin{split}
&\mathbb{E}[V(\textbf{x}(\textbf{T}, \textbf{J}))]-V(x)\\
=&\mathbb{E}[\sum_{i=0}^{\textbf{J}-1}(V(\textbf{g}(\textbf{T}_{i+1},i))-V(\textbf{x}(\textbf{T}_{i+1},i)))]\\
&+\mathbb{E}[\int_{0}^{\textbf{T}}\langle \nabla V(\textbf{x}(s,\textbf{J}_{s})), \textbf{f}(s,\textbf{J}_{s})\rangle ds]\\
&+\mathbb{E}[\int_{0}^{\textbf{T}}\frac{1}{2}\mbox{trace}(\textbf{b}(s,\textbf{J}_{s})\textbf{b}^{T}(s,\textbf{J}_{s})\nabla^{2} V(\textbf{x}(s,\textbf{J}_{s})))ds]\\
\leq&\mathbb{E}[\sum_{i=0}^{\textbf{J}-1}\Delta_{S}V(\textbf{x}(\textbf{T}_{i+1},i))+\int_{0}^{\textbf{T}} \mathcal {L}_{S}V(\textbf{x}(s,\textbf{J}_{s}))ds],
\end{split}
\end{equation}
where
\begin{equation*}\label{4.5shs:1.7}
\begin{split}
\mathcal {L}_{S}V(x):=\sup_{f\in F(x),b\in B(x)}(\langle \nabla V(x), f\rangle +\frac{1}{2}\mbox{trace}(bb^{T}\nabla^{2} V(x))
\end{split}
\end{equation*}
for all\footnotemark[2] $x\in C\cap S$,
and
\begin{equation*}\label{4.5shs:1.8}
\begin{split}
\Delta_{S}V(x):=\int_{\mathbb{R}^{m}}\sup_{g\in G(x,v)\cap S}V(g)\mu(dv)-V(x)
\end{split}
\end{equation*}
for all $x\in D\cap S$ are all well-defined and finite under Assumption 1.

\footnotetext [2]{Note that the term $\lambda(x)(\int_{\mathbb{R}^{m}}\sup_{g\in G(x,v)\cap S}V(g)\mu(dv)-V(x)))$ will be added in $\mathcal {L}_{S}V(x)$ for the case with spontaneous transitions during flows, see \cite{7.3shs}.}


\section{Stability}\label{section3}

Based on the proposed stochastic hybrid models, stability results for SHS (\ref{4.5shs:1}) are obtained via stochastic hybrid tools, such as Lyapunov-based sufficient conditions including Lagrange/Lyapunov/Lyapunov-Forster functions respectively for Lagrange/Lyapunov/asymptotic stability in probability and Foster/Lagrange-Forster functions for recurrence, (weaker) relaxed Lyapunov-based sufficient conditions including Matrosov-Foster functions and the stochastic invariance principle, as well as Lyapunov-based necessary and sufficient conditions for asymptotic stability in probability or recurrence (i.e.,converse theorems), etc. Moreover, before utilizing those techniques, the established models need to satisfy the mild regularity conditions (not necessarily $B(x)=0$), under which, a sequential compactness property for solutions is preserved leading to the nominal robustness and converse theorems when $B(x)=0$.


\subsection{Stability concepts}\label{section3.1}

Essentially, uniform global asymptotic stability in probability (UGASp) is consisting of uniform global stability in probability (UGSp) and uniform global attractivity in probability (UGAp) \cite{4.5shs}. Before showing that the certification candidate $\mathcal{U}(\xi)$ satisfying Lyapunov-based sufficient conditions is related to the notion of UGSp based on \cite[Thm. 3]{4.5shs}, the following states the definition adopted from \cite[Section VI-A]{4.5shs}.

\textbf{Definition 1.} [UGASp] The closed set $\mathcal {A}\subseteq \mathbb{R}^{n}$ is said to be UGASp if the following conditions hold.\\
\indent (a) [\emph{UGSp}]\\
   \indent\indent (a1) [\emph{Uniform Lyapunov stability in probability~(ULySp)}] If there exists $\mu>0$ such that, for each solution of $dx\in F(x)dt+B(x)d\textbf{w}$ with $x\in C\cap(\mathcal {A}+\mu \mathbb{B})$, there are no finite escape times almost surely and for any $\varepsilon, \rho>0$, there exists $\delta>0$ such that
\begin{equation}\label{4.5u}
\begin{split}
&\textbf{x}\in\mathcal {S}_{r}(\mathcal {A}+\delta \mathbb{B})\Longrightarrow\\
&\mathbb{P}(\mbox{graph}(\textbf{x})\subseteq \mathbb{R}^{2}\times(\mathcal {A}+\varepsilon \mathbb{B}))\geq1-\rho.
\end{split}
\end{equation}\\
\indent \indent (a2) [\emph{Uniform Lagrange stability in probability~(ULaSp)}] If for each solution of (\ref{4.5shs:1A}) there are no finite escape times almost surely and for any $\delta, \rho>0$ there exists $\varepsilon>0$ such that (\ref{4.5u}) holds.\\
\indent (b) [\emph{UGAp}] There are no finite escape times for (\ref{4.5shs:1A}) and for each
 $\delta>0$, $\varepsilon>0$ and $\rho>0$, there exists $\tau>0$ such that
\begin{equation}\label{4.5shs:4}
\begin{split}
&x\in\mathcal {S}_{r}(\mathcal {A}+\delta \mathbb{B})\Longrightarrow\\
&\mathbb{P}(\mbox{graph}(\textbf{x})\cap(\Gamma_{\geq \tau}\times \mathbb{R}^{2})\subset \mathbb{R}^{2}\times(\mathcal {A}+\varepsilon \mathbb{B}^{\circ}))\geq1-\rho.
\end{split}
\end{equation}

In inequality (\ref{4.5u}), the condition $\mbox{graph}(\textbf{x})\subseteq \mathbb{R}^{2}\times(\mathcal {A}+\varepsilon \mathbb{B})$ is equivalent to $\textbf{x}\in (\mathcal {A}+\varepsilon \mathbb{B})$ for all $(t,j)\in [0, \textbf{T}_{\dashv}(\omega)]\times\{0,1,\ldots,\textbf{J}_{\dashv}(\omega)-1\}$, where $(\textbf{T}_{\dashv}(\omega), \textbf{J}_{\dashv}(\omega))$ is the same with that in \cite[(4)]{4.5shs}. The set $\{\omega \in \Omega|\mbox{graph}(\textbf{x})\subseteq \mathbb{R}^{2}\times(\mathcal {A}+\varepsilon \mathbb{B})\}$ is measurable based on \cite[Thm. 14.3(i)]{Rock1998} and the measurability of $\omega\mapsto\mbox{graph}(\textbf{x}(\omega))$ \cite[Prop. 14.11 (b)\&(d)]{Rock1998}; and additionally based on the measurability of $\omega\mapsto \mbox{graph}(\textbf{x})\cap(\Gamma_{\geq \tau}\times \mathbb{R}^{2})$ \cite[Prop. 14.11 (a)]{Rock1998}, the set $\{\omega \in \Omega|\mbox{graph}(\textbf{x})\cap(\Gamma_{\geq \tau}\times \mathbb{R}^{2})\subset \mathbb{R}^{2}\times(\mathcal {A}+\varepsilon \mathbb{B}^{\circ})\}$ is measurable, where $\mbox{graph}(\textbf{x}(\omega)):=\{(t,j,z)\in\mathbb{R}_{\geq0}\times \mathbb{Z}_{\geq0}\times\mathbb{R}^{n}:(t,j)\in \mbox{dom}~\textbf{x}_{\omega}, z=\textbf{x}_{\omega}(t,j), \textbf{x}_{\omega}:=\textbf{x}(\omega)\}$. In addition, item (a2) is of the global sense, which justifies the definition for UGSp in terms of items (a1)-(a2).

Rather than directly deriving UGAp for the attractor $\mathcal {A}$, we now consider another closely related weaker stability property, i.e., uniformly global recurrence in probability (UGR) established in \cite[Thms. 1-2]{4.5shs} and \cite[Props. 3-4]{3.3shs}. As one of the weaker notions of stochastic stability properties, investigating recurrence is both practically useful and theoretically interesting \cite{4.6shs}, \cite{4.5shs}, \cite{3.2shs}. Indeed, random solutions are not necessarily prevented from reaching arbitrarily large sets with probability one even if a recurrent set exists \cite{3.5shs}. Also, in a probabilistic sense, one set is not necessarily forward invariant if it is recurrent\footnotemark[3], Therefore, recurrence provides an alternative to the Krasovskii-LaSalle invariance principle in terms of Matrosov's theorem \cite{3.3shs}, \cite{LaSalle1967} or directly based on \cite[Thm. 5]{4.5shs} with the certification candidate function.

\footnotetext [3]{Note that this is not in contradiction to the fact that a compact set which is weakly totally recurrent in
probability is also weakly forward invariant almost surely obtained in \cite[Fact 4]{5.4shs}, \cite[Coro. 4]{6.1shs}. Indeed, weak total recurrence in probability is stronger than recurrence in probability, which can be checked from their definitions in \cite[Def. 1\&(26)]{6.1shs}. }

Let us report from \cite[Section 4.1]{3.2shs} the basic definitions for recurrence relative to an open set.

\textbf{Definition 2.} [UGR] The open set $\mathcal {O}\subseteq \mathbb{R}^{n}$ is said to be UGR, if for each solution of (\ref{4.5shs:1A}) there are no finite escape times almost surely and for each $\bar{\Delta}>0$ and $\rho>0$ there exists $\tau>0$ such that
\begin{equation}\label{4.5shs:4}
\begin{split}
\textbf{x}\in \mathcal {S}_{r}(\bar{\mathcal {O}}+\bar{\Delta }\mathbb{B})\Longrightarrow\mathbb{P}(\Omega_{a}\cup\Omega_{b})\geq1-\rho,
\end{split}
\end{equation}
where $\Omega_{a}:=\{\omega\in \Omega: \mbox{graph}(\textbf{x}(\omega)) \subseteq (\Gamma_{<\tau}\times\mathbb{R}^{n})\}$
and $\Omega_{b}:=\{\omega\in \Omega: \mbox{graph}(\textbf{x}(\omega)) \cap \Gamma_{\leq\tau}\times\mathcal {O}\neq\emptyset\}$.

The above recurrence condition (\ref{4.5shs:4}) where the
probability is close to one, asks that the solution is either not
defined at time $\tau$ or else it reaches within time $\tau$. Note that the recurrence condition (\ref{4.5shs:4})
holds for initial conditions $x\in\mathcal {O}$ as shown in \cite[(8)]{3.2shs} and so it needs to be checked only
for the complement of $\mathcal {O}$ denoted as $\mathcal {O}^{c}$. Define the open set $\mathcal {O}:=\{\xi\in\tilde{\mathbb{X}}, |\xi|_{\mathcal {A}}\in [0,\delta)\cup(\Delta,+\infty)\}$ with compact set $\mathcal {O}^{c}=\{\xi\in\tilde{\mathbb{X}}, |\xi|_{\mathcal {A}}\in [\delta,\Delta]\}$ for each $0<\delta<\Delta<\infty$, where $\mathcal {A}=\{\xi\in \tilde{\mathbb{X}}|\tilde{q}=\mbox{sgn}(h)\textbf{i}, \tilde{\Theta}=0, e=0\}$. Now we present the recurrence relative to the open set $\mathcal {O}$.

Note that we drop the phrase ``relative to $\varpi$'' in the above definitions with $\varpi(x):=|x|_{\mathcal {A}}$ for all $x\in C\cup D$, see \cite{3.1shs} or \cite{4.5shs} for more details.

In the following we present two propositions for the connections between ULaSp and UGRp, as well as the connections between UGASp and UGR, which are borrowed from \cite[Prop. 2.2\&2.4]{3.1shs}, respectively. In what follows, the results on UGASp and UGR are based on the following two propositions, besides the definition of UGASp, i.e., ULaSp, ULySp and UGAp.

\textbf{Proposition 1}.
The compact set $\mathcal {A}\subseteq \mathbb{R}^{n}$ is UGASp if it is UGSp and for each $0<\delta<\Delta<\infty$,
the open set $\{x\in\mathbb{R}^{n}, |x|_{\mathcal {A}}\in [0,\delta)\cup(\Delta,+\infty)\}$ is UGRp.

\textbf{Proposition 2}.The open set $\mathcal {O}\subseteq \mathbb{R}^{n}$ is UGRp if its closure is uniformly Lagrange
stable in probability conditioned on not reaching $O$ and, for
each $\Delta > 0$, the set $O\cup(\mathbb{R}^{n}\setminus O+\Delta \mathbb{B})$ is UGRp.

The next result is used to establish uniform global
recurrence, i.e., uniform global recurrence of
open set $\mathcal {O}\subseteq \mathbb{R}^{n}$ for $\mathcal {H}$ is equivalent to uniform global recurrence of $\mathcal {O}\subseteq \mathbb{R}^{n}$
for the truncated data $\mathcal {H}_{\cap}$, where $\mathcal {H}_{\cap}= (C\cap \mathcal {O}^{c}, F,D\cap \mathcal {O}^{c},G\cap \mathcal {O}^{c}, \mu)$.

\textbf{Proposition 3}.
Let $\varpi: C \cup D \rightarrow \overline{\mathbb{R}_{\geq0}}$ and suppose there are no finite escape
times for (\ref{4.5shs:1A}). The open set $\mathcal {O}\subseteq \mathbb{R}^{n}$ is UGR relative to $\varpi$ for $\mathcal {H}$ if and only if it is UGR relative to $\varpi$ for $\mathcal {H}_{\cap}$.

As shown in \cite{3.1shs}, since we do not necessarily assume that the open/closed set is bounded, we ask for some uniformity in
probability over solutions from certain sets of initial conditions. As a result, sometimes we drop the word ``uniform'' for the compact set $\mathcal {A}$ or the open, bounded set $\mathcal {O}$ in the definitions of stability properties, as ULaSp, ULySp, UGAp, UGR and UGASp, see \cite{3.7shs}.


\subsection{Sufficient conditions}\label{section3.2}

Now we adopt these Lyapunov functions proposed in \cite{3.1shs} to analyze stability in probability.



A certification candidate for $\mathcal {H}$ (\ref{4.5shs:1}) is a Lagrange
function\footnotemark[4] for $\mathcal {A}$ relative to $\varpi$ for (\ref{4.5shs:1})
if there exist $\alpha_{1}, \alpha_{2}\in \mathcal {G}_{\infty}$ such that
\begin{subequations}\label{th1:b}
\begin{align}
&\alpha_{1}(|x|_{\mathcal {A}})\leq V(x), ~x\in C\cup D\cup G(D\times \mathcal {V}),\label{th1:A}\\
&V(x)\leq \alpha_{2}(\varpi(x)),~x\in C\cup D,\label{th1:B}\\
&V(\phi(t))\leq V(x), ~x\in C, ~\phi \in \mathcal {S}_{C,F}(x), ~t\in \hbox{dom}(\phi),\label{th1:C}\\
&\int_{\mathbb{R}^{m}}\sup_{g\in G(x,v)}V(g)\mu(dv) \leq V(x), x\in D.\label{th1:D}
\end{align}
\end{subequations}
In addition, it is a Lyapunov function if the above conditions hold with $\alpha_{1}, \alpha_{2}\in \mathcal {K}_{\infty}$.

\footnotetext [4]{A partially Lipschitz certification candidate for $\mathcal {H}$ (\ref{4.5shs:1}) is a
regular Lagrange (Lyapunov) function for $\mathcal {A}$ relative to $\varpi$ for (\ref{4.5shs:1}) if it satisfies
the conditions (\ref{th1:A}), (\ref{th1:B}), and (\ref{th1:D}) and $V^{\circ}(x,F)\leq 0$, $x\in C\setminus L_{V}(0)$, $f\in F(x)$.}

\textbf{Theorem 1} (Lagrange/Lyapunov function implies Lagrange/Lyapunov stability) \cite[Thm. 4.1\&4.2]{3.1shs}.
Let $\mathcal {A}\subset \mathbb{R}^{n}$ be closed and $\varpi: C\cup D\rightarrow \overline{\mathbb{R}_{\geq0}}$. Suppose Assumption 1 holds and that either $\mathcal {A}$ is bounded or there are no finite escape times
for (\ref{4.5shs:1A}). Under these conditions, if there exists a Lagrange/Lyapunov function for
$\mathcal {A}$ relative to $\varpi$ for (\ref{4.5shs:1}) then $\mathcal {A}$ is ULaSp/ULySp relative to $\varpi$ for (\ref{4.5shs:1}).

Define $\mathcal {H}_{\cap}:=(C\cap \mathcal {O}^{c}, (F,B),D\cap \mathcal {O}^{c},G\cap \mathcal {O}^{c}, \lambda, \mu)$.

A certification candidate for $\mathcal {H}_{\cap}$ is called a Foster function for $\mathcal {O}$ with relative to $\varpi$ for (\ref{4.5shs:1}) if there exist $\rho>0$ and $\alpha_{2}\in \mathcal {G}_{\infty}$ and such that
\begin{subequations}\label{th2:b}
\begin{align}
&V(x)\leq \alpha_{2}(\varpi(x)),~x\in C_{\cap}\cup D_{\cap},\label{th2:B}\\
&V(\phi(t))\leq V(x)-\rho t, ~x\in C_{\cap}, ~\phi \in \mathcal {S}_{C_{\cap},F}(x), ~t\in \hbox{dom}(\phi),\label{th2:C}\\
&\int_{\mathbb{R}^{m}}\sup_{g\in G_{\cap}(x,v)}V(g)\mu(dv) \leq V(x)-\rho, x\in D_{\cap}.\label{th2:D}
\end{align}
\end{subequations}
In addition, a partially Lipschitz certification candidate for $\mathcal {H}_{\cap}$ is a
regular Foster function for $\mathcal {O}$ relative to $\varpi$ for (\ref{4.5shs:1}) if it satisfies
the conditions (\ref{th2:B}) and (\ref{th2:D}) and $V^{\circ}(x,F)\leq 0$, $x\in C_{\cap}\setminus L_{V}(0)$, $f\in F(x)$.

Note that the expression in (\ref{th1:C}) and (\ref{th2:C}) are very general, in order to respond to the spontaneous transition during flows.
Even if certification candidate $V$ is continuous on $C$, the spontaneous transition of $x$ during $C$ renders that we could use the Clarke generalized directional derivative of $V$, i.e., the regular certification candidate $V$, which implies (\ref{th1:C}) according to \cite[Props. 4.1\&4.2\&4.3]{3.1shs}, respectively for Lagrange function, Lyapunov function and Foster function.

\textbf{Theorem 2} (Foster function implies recurrence) \cite[Thm. 4.3]{3.1shs}.
Let $\mathcal {O}\subset \mathbb{R}^{n}$ be open and $\varpi: C\cup D\rightarrow \overline{\mathbb{R}_{\geq0}}$. Suppose Assumption 1 holds and that either $\mathcal {A}$ is bounded or there are no finite escape times
for (\ref{4.5shs:1A}). Under these conditions, if there exists a Foster function for
$\mathcal {O}$ relative to $\varpi$ for (\ref{4.5shs:1}) then $\mathcal {O}$ is UGR relative to $\varpi$ for (\ref{4.5shs:1}).

A partially Lipschitz (respectively, partially $C^{1}$) certification candidate for $\mathcal {H}_{\cap}$ is called a regular
(respectively, partially $C^{1}$) Lagrange-Foster function for $\mathcal {O}$ with respect to $\varpi$ for (\ref{4.5shs:1}) if there exist $\alpha_{1}, \alpha_{2}\in \mathcal {G}_{\infty}$ and $\rho\in \mathcal {P}\mathcal {D}(\mathcal {O})$ such that
\begin{subequations}\label{th3:b}
\begin{align}
&\alpha_{1}(|x|_{\mathcal {A}})\leq V(x), ~x\in C_{\cap}\cup D_{\cap}\cup G_{\cap}(D_{\cap}\times \mathcal {V}),\label{th3:A}\\
&V(x)\leq \alpha_{2}(\varpi(x)),~x\in C_{\cap}\cup D_{\cap},\label{th3:B}\\
&V^{\circ}(x,F)\leq-\rho(x), ~x\in C_{\cap}\setminus L_{V}(0),~f\in F(x),\label{th3:C}\\
&\int_{\mathbb{R}^{m}}\sup_{g\in G_{\cap}(x,v)}V(g)\mu(dv) \leq V(x)-\rho(x), x\in D_{\cap}.\label{th3:D}
\end{align}
\end{subequations}

\textbf{Theorem 3} (Lagrange-Foster function implies recurrence) \cite[Thm. 4.4]{3.1shs}.
Let $\mathcal {O}\subset \mathbb{R}^{n}$ be open and $\varpi: C\cup D\rightarrow \overline{\mathbb{R}_{\geq0}}$. Suppose Assumption 1 holds and that either $\mathcal {O}$ is bounded or there are no finite escape times
for (\ref{4.5shs:1A}). Under these conditions, if there exists a regular Lagrange-Foster function for $\mathcal {O}$ relative to $\varpi$ for (\ref{4.5shs:1}) then $\mathcal {O}$ is UGR relative to $\varpi$ for (\ref{4.5shs:1})

A partially Lipschitz (respectively, partially $C^{1}$) certification candidate for $\mathcal {H}$ is called a regular
(respectively, partially $C^{1}$) Lyapunov-Foster function for $\mathcal {A}$ with respect to $\varpi$ for (\ref{4.5shs:1}) if there exist $\alpha_{1}, \alpha_{2}\in \mathcal {K}_{\infty}$ and $\rho\in \mathcal {P}\mathcal {D}(\mathcal {A})$ such that
\begin{subequations}\label{th4:b}
\begin{align}
&\alpha_{1}(|x|_{\mathcal {A}})\leq V(x), ~x\in C\cup D\cup G(D\times \mathcal {V}),\label{th4:A}\\
&V(x)\leq \alpha_{2}(\varpi(x)),~x\in C\cup D,\label{th4:B}\\
&V^{\circ}(x,F)\leq-\rho(x), ~x\in C\setminus L_{V}(0),~f\in F(x),\label{th4:C}\\
&\int_{\mathbb{R}^{m}}\sup_{g\in G(x,v)}V(g)\mu(dv) \leq V(x)-\rho(x), x\in D.\label{th4:D}
\end{align}
\end{subequations}

\textbf{Theorem 4} (Lyapunov-Foster function implies UGASp) \cite[Thm. 4.5]{3.1shs}.
Let $\mathcal {A}\subset \mathbb{R}^{n}$ be closed, $\varpi: C\cup D\rightarrow \overline{\mathbb{R}_{\geq0}}$ be lower semicontinuous. Suppose that
\begin{equation}\label{th1:a}
\begin{split}
&\forall\delta_{\varpi}>0, \exists\delta_{\mathcal {A}}>0:\\
&((\mathcal {A}+\delta_{\mathcal {A}}\mathbb{B}^{\circ})\cap (C\cup D))\subset\{x\in C\cup D: \varpi(x)<\delta_{\varpi}\},
\end{split}
\end{equation}
Assumption 1 holds, and that either A is bounded or there are no
finite escape times for (\ref{4.5shs:1A}). Under these conditions, if there exists a regular Lyapunov-Foster function for $\mathcal {A}$ relative to $\varpi$ for (\ref{4.5shs:1}), then $\mathcal {A}$ is UGASp relative to $\varpi$ for (\ref{4.5shs:1}).

Note that the phrase relative to ``$\varpi$'' is dropped in Definition 1 since $\varpi(x):=|x|_{\mathcal {A}}$ for all $x\in C\cup D$. In this case, the mapping $\varpi(x)$ is continuous as it is the Euclidean distance to
the closed set $\mathcal {A}$.

Note that in the above conditions (\ref{th1:b}), (\ref{th2:b}), (\ref{th3:b}), (\ref{th4:b}), there are two kinds of functions $\mathcal {K}_{\infty}\subset\mathcal {G}_{\infty}$ for the radial unboundedness of those certification candidates, i.e., $\mathcal {G}_{\infty}$ for Lagrange function (\ref{th1:b}), Foster function (\ref{th2:b}) and Lagrange-Foster function (\ref{th3:b}), while $\mathcal {K}_{\infty}$ for Lyapunov function (\ref{th1:b}) and Lyapunov-Foster function (\ref{th4:b}). Based on the definitions of $\mathcal {K}_{\infty}, \mathcal {G}_{\infty}: \mathbb{R}_{\geq0}\rightarrow \mathbb{R}_{\geq0}$, we know that $\mathcal {K}_{\infty}$-class function is continuous, strictly increasing, zero at zero and unbounded, while $\mathcal {G}_{\infty}$-class function is continuous, nondecreasing and unbounded. So it is easy to check that \\
\indent 1) $\mathcal {K}_{\infty}$-class function in (\ref{th1:b}) and (\ref{th4:b}) implies $V(x)=0$ iff $x\in \mathcal {A}$ when we set $\varpi(x):=|x|_{\mathcal {A}}$ for all $x\in C\cup D$, as shown in \cite[Thm. 8]{6.1shs}, however, $V(x)=0$ \textcolor{red}{may hold not only for} if $x\in \mathcal {A}$ for $\mathcal {G}_{\infty}$-class function; \\
\indent 2) the radial unboundedness with $\mathcal {K}_{\infty}$-class function is much stronger than that with $\mathcal {G}_{\infty}$-class function, since $\mathcal {K}_{\infty}$-class function is strictly increasing;\\
\indent 3) for ULySp or UGASp in (\ref{th1:b}) and (\ref{th4:b}), we have to apply $\mathcal {K}_{\infty}$-class function for the radial unboundedness, ULySp (\ref{4.5u}) as well as UGASp says that ``no matter how close we want the solutions to remain to $\mathcal {A}$ (within a distance $\varepsilon>0$) with probability, $\cdots$''\cite{5.2shs}, so that we have to stipulate that $V(x)=0$ if and only if $x\in \mathcal {A}$. Otherwise, if we apply $\mathcal {G}_{\infty}$-class function, it means that $V(x)=0$ when $x\in \mathcal {A}$ or $|x|_{\mathcal {A}}=c$ \textcolor{red}{(e.g., $\alpha_{1}(|x|_{\mathcal {A}})=\alpha_{1}(s)\equiv0 for s\in [0,c])$} for certain $c>0$, then $x$ stays in
$(\mathcal {A}+c\mathbb{B})$, so it is not possible to let the solutions remain sufficiently close to $\mathcal {A}$, i.e., $\varepsilon$ has to be larger than or equal to $c$. However, for ULaSp or UGR in (\ref{th1:b}), (\ref{th2:b}) and (\ref{th3:b}), we do not have to do like this and the mild $\mathcal {G}_{\infty}$-class function is enough, since ULaSp or UGR is about ``no matter how far from $\mathcal {A}$ or the closure of $\mathcal {O}$ the solutions start (within a distance $\delta>0$), $\cdots$''\cite{5.2shs}.


\subsection{Relaxed sufficient conditions}\label{section3.3}

As analyzed in \cite[Sec. 7]{4.1shs}, for the case of compact attractors, relaxed sufficient conditions include Matrosov-Foster functions \cite{3.1shs}, \cite{3.5shs}, \cite{5.2shs} and the stochastic invariance principle \cite{5.4shs}, \cite{6.1shs}, \cite{4.2shs}, \cite{4.4shs}.

\subsubsection{Matrosov-Foster functions for UGR\&UGASp}\label{section3.3.1}



\begin{figure*}[!htb]
\setlength{\unitlength}{0.7mm}
$$\begin{picture}(200,66)
\put(0,54){$u_{c,0}=0$}
\put(19,55.5){\vector(1,0){5}}
\put(25,54){$u_{c,1}\leq0$}
\put(48,54){$\hbox{or}$}
\put(44,55.5){\line(0,1){7}}
\put(44,55.5){\line(0,-1){7}}
\put(44,62.5){\line(1,0){2}}
\put(44,48.5){\line(1,0){2}}
\put(46.7,60.9){$<0$,}
\put(46.7,47.8){$=0$}
\put(56.5,61){($u_{c,2}>0$~\hbox{or}~$\leq0$;~$\hbox{stop};~m=1$);}
\put(56.5,49.3){\vector(1,0){5}}

\put(63,48.8){$u_{c,2}\leq0$}
\put(86,48.3){$\hbox{or}$}
\put(82,49.3){\line(0,1){7}}
\put(82,49.3){\line(0,-1){7}}
\put(82,56.3){\line(1,0){2}}
\put(82,42.3){\line(1,0){2}}
\put(84.7,54.7){$<0$,}
\put(84.7,41.6){$=0$}
\put(94.5,54.7){($u_{c,3}>0$~\hbox{or}~$\leq0$;~$\hbox{stop};~m=2$);}
\put(94.5,43.3){\vector(1,0){5}}

\put(101,42.8){$u_{c,3}\leq0$}
\put(124,42.3){$\hbox{or}$}
\put(120,43.3){\line(0,1){7}}
\put(120,43.3){\line(0,-1){7}}
\put(120,50.3){\line(1,0){2}}
\put(120,36.3){\line(1,0){2}}
\put(122.7,48.7){$<0$,}
\put(122.7,35.6){$=0$}
\put(132.5,48.7){($u_{c,4}>0$~\hbox{or}~$\leq0$;~$\hbox{stop};~m=3$);}
\put(132.5,37.3){\vector(1,0){5}}

\put(139,36.8){$u_{c,5}\leq0,~\cdots$}

%
%
%
%

\put(-7,9.45){$u_{d,0}\leq0$}
\put(16,9.45){$\hbox{or}$}
\put(12,10.95){\line(0,1){9.65}}
\put(12,10.95){\line(0,-1){9.65}}
\put(12,20.6){\line(1,0){2}}
\put(12,1.3){\line(1,0){2}}
\put(14.7,19){$=0$}
\put(14.7,0.6){$<0$,}
\put(24.5,0.7){($u_{d,1}>0$~\hbox{or}~$\leq0$;~$\hbox{stop};~m=0$);}
\put(25,20.5){\vector(1,0){5}}

\put(33,19){$u_{d,1}\leq0$}
\put(56,19){$\hbox{or}$}
\put(52,20.5){\line(0,1){7}}
\put(52,20.5){\line(0,-1){7}}
\put(52,27.5){\line(1,0){2}}
\put(52,13.5){\line(1,0){2}}
\put(54.7,25.9){$<0$,}
\put(54.7,12.8){$=0$}
\put(64.5,26){($u_{d,2}>0$~\hbox{or}~$\leq0$;~$\hbox{stop};~m=1$);}
\put(64.5,14.3){\vector(1,0){5}}

\put(71,13.8){$u_{d,2}\leq0$}
\put(94,13.3){$\hbox{or}$}
\put(90,14.3){\line(0,1){7}}
\put(90,14.3){\line(0,-1){7}}
\put(90,21.3){\line(1,0){2}}
\put(90,7.3){\line(1,0){2}}
\put(92.7,19.7){$<0$,}
\put(92.7,6.6){$=0$}
\put(102.5,19.7){($u_{d,3}>0$~\hbox{or}~$\leq0$;~$\hbox{stop};~m=2$);}
\put(102.5,8.3){\vector(1,0){5}}

\put(109,7.8){$u_{d,3}\leq0$}
\put(132,7.3){$\hbox{or}$}
\put(128,8.3){\line(0,1){7}}
\put(128,8.3){\line(0,-1){7}}
\put(128,15.3){\line(1,0){2}}
\put(128,1.3){\line(1,0){2}}
\put(130.7,13.7){$<0$,}
\put(130.7,0.6){$=0$}
\put(140.5,13.7){($u_{d,4}>0$~\hbox{or}~$\leq0$;~$\hbox{stop};~m=3$);}
\put(140.5,2.3){\vector(1,0){5}}

\put(147,1.8){$u_{d,5}\leq0,~\cdots$}
\end{picture}$$
\caption{Relationship among $u_{c,i}$ and $u_{d,i}$ of a family of Matrosov-Foster functions. $u_{c,i}$ and $u_{d,i}$ share the same number $m$ which is negotiated by the condition $j\in\{0,1,\ldots,m\}$ in item 2c). More spcifically, once $u_{c,i_{0}}<0$ holds for certain $i_{0}$, then it stops at $m=i_{0}$ since $u_{c,i+1}>0$ may appear; while once $u_{d,j_{0}}<0$ holds for certain $j_{0}$, then it stops at $m=j_{0}$ since $u_{d,i+1}>0$ may appear; and $m:=\min\{i_{0}, j_{0}\}$, where item 2c) implies that the constructed Marosov-Foster functions are such that $|i_{0}-j_{0}|\leq1$. For example in the stochastic bouncing ball \cite[Prop. 6.1]{3.1shs}, one obtains $u_{c,0}=0$ and $u_{d,0}<0$, while $u_{c,1}<0$ and $u_{d,1}>0$, with $i_{0}=1$ and $j_{0}=0$, thus $m=\min\{i_{0}, j_{0}\}=0$.}\label{matro}
\end{figure*}

Sometimes finding a single certification candidate
satisfying the conditions for UGASp is not easy. In this case, searching for a family
of functions is useful as shown in \cite{3.1shs}, \cite{3.5shs}, \cite{5.2shs} for ensuring UGR, where none of which satisfies the conditions presented in the previous sections, but if taken together, a strong conclusion about asymptotic stability is still permitted.

Given an open set $\mathcal {O}\subset \mathbb{R}^{n}$ whose complement is compact. Let $m\in \mathbb{Z}_{\geq1}$. Following \cite[\S5.2]{5.2shs} for $\mathcal {H}= (C, (F,B),D,G, \mu)$ and \cite[Sec. 5.2]{3.1shs} for $\mathcal {H}= (C, F,D,G, \mu)$, we say that the $m+1$ functions $V_{i}$, $i\in\{0,1,\ldots,m\}$ are \emph{Matrosov-Foster functions} for $\mathcal {O}$ if there exist continuous functions $u_{c,i}: C\cap \mathcal {O}^{c}\rightarrow \mathbb{R}$ and
$u_{d,i} : D\cap \mathcal {O}^{c}\rightarrow \mathbb{R}$, $i\in \{0,1,\ldots,m\}$, such that
\begin{itemize}
  \item [1)] for all $i\in \{0,1,\ldots,m\}$, $V_{i}$ is a certification candidate for
$(C\cap \mathcal {O}^{c}, (F,B), D\cap \mathcal {O}^{c}, G\cap \mathcal {O}^{c}, \mu)$ that is bounded
on $(C\cup D)\cap \mathcal {O}^{c}$, and\footnotemark[5]
\begin{subequations}\label{1shs:1}
\begin{align}
\mathcal {L}_{\mathcal {O}^{c}}V_{i}(x)\leq& u_{c,i}(x),~\forall x\in C\cap \mathcal {O}^{c},\label{1shs:1A}\\
\Delta_{\mathcal {O}^{c}}V_{i}(x)\leq& u_{d,i}(x),~\forall x\in D\cap \mathcal {O}^{c},\label{1shs:1B}
\end{align}
\end{subequations}
  \item [2)]
\begin{itemize}
  \item [2a)] $u_{c,0}(x)\leq0$ for all $x\in C\cap \mathcal {O}^{c}$,
  \item [2b)] $u_{d,0}(x)\leq0$ for all $x\in D\cap \mathcal {O}^{c}$,
  \item [2c)] with the functions $u_{c,m+1},u_{d,m+1}: \mathbb{R}^{n}\rightarrow \{1\}$,
for each $j\in\{0,1,\ldots,m\}$ we have
\begin{itemize}
  \item [i)] if $x\in C\cap \mathcal {O}^{c}$ and $u_{c,i}(x)=0$ for all $i\in\{0,1,\ldots,j\}$ then $u_{c,j+1}(x)\leq0$,
  \item [ii)] if $x\in D\cap \mathcal {O}^{c}$ and $u_{d,i}(x)=0$ for all $i\in\{0,1,\ldots,j\}$ then $u_{d,j+1}(x)\leq0$.
\end{itemize}
\end{itemize}
\end{itemize}

As pointed out in \cite[pg. 3131]{5.2shs} or \cite[pg. 10]{3.1shs}, item 2) amounts to a nested negative
semi-definite condition on functions $u_{c,i}$ and $u_{d,i}$. Items 2a)-2b) require that functions $u_{c,0}$
and $u_{d,0}$ are never positive; while item 2c) enforces that $u_{c,j+1}$ ($u_{d,j+1}$) can be positive only when at least one of the functions $u_{c,i}$ ($u_{d,i}$) is negative for $i\in\{0,1,\ldots,j\}$. The definition of $u_{c,m+1}$ (or $u_{d,m+1}$) in item 2c) also implies that there are no points in $C\cap \mathcal {O}^{c}$ (or $D\cap \mathcal {O}^{c}$) such that all of $u_{c,i}$ (or $u_{d,i}$) are zero for each $i\in\{0,1,\ldots,m\}$. See Fig. \ref{matro} for the evolutional relationship among $u_{c,i}$ or $u_{d,i}$ of a family of Matrosov-Foster functions.

\footnotetext [5]{Note that \cite[(29)]{3.1shs} presents a more general form for $\mathcal {H}= (C, F,D,G, \mu)$, i.e., $V_{0}(\phi(t))\leq V_{0}(x)$ and $V^{\circ}_{i}(x;f)\leq u_{c,i}(x)$ for $x\in C_{\cap}$, $\phi \in \mathcal {S}_{C_{\cap},F}(x)$, $f\in F(x)$, $t\in \hbox{dom}(\phi)$ and $i\in\{1,2,\ldots,m\}$. The two kinds of form are essentially the same, since generalized directional directive is an upper bound for the usual directional derivative \cite[pg. 99]{2015frame[30]}.}

Sometimes $u_{c,0}(x)\leq0$ is written as $u_{c,0}(x)=0$, as \cite[Sec. 5.2]{3.1shs}. This is because for UGASp or UGR, the precondition for using (relaxed) Matrosov-Foster functions is UGSp or LaSp as shown in Coro. 6\&7, i.e., the first one of a family of certification candidates should satisfy $V^{\circ}_{0}(x;f)=0$ or $\int_{\mathbb{R}^{m}}\sup_{g\in G(x,v)}V_{0}(g)\mu(dv)= V_{0}(x)$ (i.e., UGSp or LaSp as shown in (\ref{th1:b})). If the fist certification candidate satisfies both of $V^{\circ}_{0}(x;f)<0$ and $\int_{\mathbb{R}^{m}}\sup_{g\in G(x,v)}V_{0}(g)\mu(dv< V_{0}(x)$ (i.e., satisfying (\ref{th4:b})), then UGASp ensues, so there is only one single function in a family of Matrosov-Foster functions, which is not wrong but defeats the purpose of (relaxed) Matrosov-Foster functions. In other words, at least one of $u_{c,0}(x)$ and $u_{d,0}(x)$ should be zero, when using Matrosov-Foster functions or stochastic invariance principle, as \cite[Exm. 1]{6.1shs}, \cite[Exm. 4]{4.2shs}, \cite[Exm. A]{4.4shs} or \cite[Sec. 6.1]{3.1shs}.

\textbf{Theorem 5} \cite[Thm. 5.2]{3.1shs}.
Let $\mathcal {O}\subset \mathbb{R}^{n}$ be open and such that $\mathbb{R}^{n}\setminus \mathcal {O}$ is compact.
If there exist Matrosov-Foster functions for $\mathcal {O}$ then there exists a Foster
function for $\mathcal {O}$.

According to Thm. 2 where Foster function implies recurrence, Thm. 5 shows that Matrosov-Foster functions for $\mathcal {O}$ imply recurrence of $\mathcal {O}$. Following Prop. 1\&2, we obtain the following corollaries.

\textbf{Corollary 6} (UGR) \cite[Coro. 5.3]{3.1shs}, \cite[Thm. 13]{5.2shs}.
Let $\mathcal {O}\subset \mathbb{R}^{n}$ be open and bounded. For
the system (\ref{4.5shs:1}), suppose Assumption 1 holds and that $\mathcal {O}$
is ULaSp conditioned on stopping at $\mathcal {O}$. If for each $\Delta>0$ there exist Matrosov-Foster functions for $\mathcal {O}\cup(\mathbb{R}^{n}\setminus (\bar{\mathcal {O}}+\Delta\mathbb{B}))$ for (\ref{4.5shs:1}) then $\mathcal {O}$ is UGR for (\ref{4.5shs:1}).

\textbf{Corollary 7} (UGASp) \cite[Coro. 5.4]{3.1shs}, \cite[Thm. 14]{5.2shs}.
Let $\mathcal {A}\subset \mathbb{R}^{n}$ be compact. Suppose Assumption
1 holds and that $\mathcal {A}$ is UGSp for (\ref{4.5shs:1}). If for each pair of positive real
numbers $(\delta, \Delta)$, there exist Matrosov-Foster functions for
$(\mathcal {A}+\delta \mathbb{B}^{\circ})\cup(\mathbb{R}^{n}\setminus (\mathcal {A}+\Delta\mathbb{B}))$ for (\ref{4.5shs:1}) then $\mathcal {A}$ is uniformly globally asymptotically stable in probability for (\ref{4.5shs:1}).

\subsubsection{Stochastic invariance principle for UGASp\&UGR}\label{section3.3.2}

For SHIs satisfying mild regularity conditions (not necessarily the Feller property), when a Lyapunov function can be found that establishes Lyapunov and Lagrange stability in probability, UGASp ensues when there are no almost surely complete solutions
remains almost surely in a non-zero level set of the Lyapunov function, as shown in \cite{5.4shs}, \cite{6.1shs}, \cite{4.2shs} and \cite{4.4shs}.

Now we present the results in \cite{6.1shs} based on stochastic invariance principle for UGASp. Let $K_{\delta, \Delta}:=\{x\in \mathbb{R}^{n}: |x|_{\mathcal {A}}\in[\delta, \Delta]\}$.

\textbf{Theorem 8} \cite[Thm. 6]{6.1shs}.
Suppose the compact set $\mathcal {A}$ is UGSp for SDDIs in (\ref{4.5shs:1}).
The set  $\mathcal {A}$ is UGASp for SDDIs in (\ref{4.5shs:1}) if for each $0<\delta<\Delta<\infty$,
the set $K_{\delta, \Delta}$ contains no compact set that is almost
surely weakly forward invariant for SDDIs in (\ref{4.5shs:1}).

\textbf{Theorem 9} \cite[Thm. 8]{6.1shs}.
Let $V$ be a Lyapunov function relative to a compact set $\mathcal {A}\subset \mathbb{R}^{n}$ for SDDIs in (\ref{4.5shs:1}). Then, $\mathcal {A}$ is UGASp if and only if there does not exist an almost surely complete solution $x$
that remains in a non-zero level set of the Lyapunov function almost surely.

Next we present the similar results for UGR.

\textbf{Theorem 10} \cite[Thm. 7]{6.1shs}.
Suppose the compact set $\bar{\mathcal {O}}$ is ULaSp for SDDIs $\mathcal {H}_{\cap}=(C\cap \mathcal {O}^{c}, F, D\cap \mathcal {O}^{c},G\cap \mathcal {O}^{c}, \lambda, \mu))$ in (\ref{4.5shs:1}). Then,
$\mathcal {O}$ is UGR for SDDIs $\mathcal {H}_{\cap}$ if for each $\Delta>0$,
there does not exist an almost surely weakly forward invariant
set contained in the compact set $(O+\Delta \mathbb{B})\setminus \mathcal {O}$ for SDDIs $\mathcal {H}_{\cap}$.

\textbf{Theorem 11} \cite[Thm. 9]{6.1shs}.
Let $V$ be a Foster function relative to an open,
bounded set $\mathcal {O}\subset \mathbb{R}^{n}$ for SDDIs $\mathcal {H}$.
Then, $\mathcal {O}$ is UGR if and only if there does not exist an almost
surely complete solution $x$ that remains almost surely in the set
$L_{V}(c)\cap (\mathbb{R}^{n} \setminus \mathcal {O})$ for every $c\geq0$ for which $L_{V}(c)\cap (\mathbb{R}^{n} \setminus \mathcal {O})$ is non-empty.

Recurrence principles and the stochastic invariance principle are weak sufficient conditions for stability properties in probability \cite[Chap. 8]{2012teel}, \cite{5.4shs}, \cite{6.1shs}, \cite{6.1shs}, as shown in Fig. \ref{ttfram}, where `weak' means that Lyapunov functions satisfy non-increasing conditions along solutions (as opposed to strong Lyapunov functions whose derivatives along the system solutions are strictly negative outside the attractor). Unlike the invariance principles \cite{LaSalle1967}, which can be used to assert convergence of random solutions for time-invariant systems as introduced in \cite[Introduction]{6.1shs}, UGRp can be applied to assert the UGAp of the attractor for SHS $\mathcal {H}$ (\ref{4.5shs:1}) as shown in Prop. 1, essentially from the point of view of the Matrosov's theorem \cite[Section X-E]{5.2shs} or \cite[Sec. 5.2]{3.1shs}. Indeed, the sufficient recurrence-Lyapunov function in \cite[Thm. 5]{4.5shs} obviously satisfies the Matrosov property presented in \cite[Section X-E]{5.2shs}, which thus connects UGAp to UGRp in terms of \cite[Thm. 1]{4.5shs}. Compared with the invariance principle, Matrosov's theorem does not necessarily require the knowledge about the system solutions to analyze the asymptotically convergence properties, as analyzed in \cite{3.3shs} and \cite{3.5shs} for the time-varying, stochastic difference inclusions. When specialized to time-invariant or almost periodic cases, Matrosov properties provide an alternative to the invariance principle for deterministic SHSs \cite[Chapter. 8.2]{2012teel} or stochastic systems \cite[Chapter 9]{yin2010}. The readers are referred to \cite[Section IV]{5.4shs} and \cite[Section VIII]{6.1shs} for comparisons between invariance principle and recurrence. See Fig. \ref{ttfram} for a summary of results on stability properties.

\begin{figure*}[!htb]
\setlength{\unitlength}{0.7mm}
$$\begin{picture}(260,200)
\put(7,181){\vector(0,-1){121}}
\put(6,181){\line(1,0){2}}

\put(53,181){\vector(0,-1){41}}
\put(52,181){\line(1,0){2}}
\put(59,181){\vector(0,-1){41}}
\put(58,181){\line(1,0){2}}
\put(60,157){Definition}

\put(4,180){\line(0,1){20}}
\put(4,180){\line(1,0){156}}
\put(4,200){\line(1,0){156}}
\put(160,180){\line(0,1){20}}
\put(17,191){ULaSp for $\mathcal {A}$:}
\put(16.5,184){Laf (\ref{th1:b}) for $\mathcal {H}$}
\put(160,190){\vector(1,0){40}}
\put(170,191){Definition}

\multiput(56,180)(0,1){20}{\line(0,1){0.5}}
\multiput(108,180)(0,1){20}{\line(0,1){0.5}}
\put(69,191){ULySp for $\mathcal {A}$:}
\put(69,184){Lyf (\ref{th1:b}) for $\mathcal {H}$}
\put(121,191){UGAp for $\mathcal {A}$:}
\put(130,184){for $\mathcal {H}$}

\put(100,155){\line(0,1){20}}
\put(100,155){\line(1,0){60}}
\put(100,175){\line(1,0){60}}
\put(160,155){\line(0,1){20}}
\put(115,166){LyF (\ref{th4:b}) for $\mathcal {H}$:}
\put(124,159){for $\mathcal {A}$}
\multiput(162,166)(1,0){38}{\line(1,0){0.5}}
\put(162,166){\vector(-1,0){2}}
\put(160,164){\vector(1,0){40}}
\put(172,167){Thm. 17}
\put(166.5,160){Thm. 4~or~17}

\put(100,130){\line(0,1){20}}
\put(100,130){\line(1,0){60}}
\put(100,150){\line(1,0){60}}
\put(160,130){\line(0,1){20}}
\put(118,141){Stochastic IP}
\put(125,134){for $\mathcal {H}$}
\multiput(162,141)(1,0){38}{\line(1,0){0.5}}
\put(162,141){\vector(-1,0){2}}
\put(160,139){\vector(1,0){40}}
\put(172,142){Thm. 9}
\put(167.5,134.7){Thm. 8~or~9}

\put(26.3,115){\line(0,1){25}}
\put(26.3,115){\line(1,0){60}}
\put(26.3,140){\line(1,0){60}}
\put(86.3,115){\line(0,1){25}}
\put(43.5,133){UGSp for $\mathcal {A}$:}
\put(44,126){LaSp\&LySp}
\put(51,119){for $\mathcal {H}$}
\put(86.3,127.5){\line(1,0){9.6}}

\put(96,140){\line(1,0){4}}
\put(96,115){\line(1,0){4}}
\put(96,115){\line(0,1){25}}
\put(89.3,126.2){$\oplus$}

\put(100,105){\line(0,1){20}}
\put(100,105){\line(1,0){60}}
\put(100,125){\line(1,0){60}}
\put(160,105){\line(0,1){20}}
\put(124.5,116){UGR:}
\put(121,109){$\mathcal {O}_{s}$ for $\mathcal {H}$}
\put(160,115){\vector(1,0){40}}
\put(173,116){Prop. 1}

\put(200,105){\line(0,1){95}}
\put(200,105){\line(1,0){40}}
\put(200,200){\line(1,0){40}}
\put(240,105){\line(0,1){95}}
\put(200.5,156){(Nominally robust)}
\put(209,149){UGASp: $\mathcal {A}$}
\put(215,142){for $\mathcal {H}$}

\put(130,99){\vector(0,1){8}}
\put(130,99){\vector(0,-1){2}}
\put(132,99){Prop. 3}

\put(10,77){\line(0,1){20}}
\put(10,77){\line(1,0){60}}
\put(10,97){\line(1,0){60}}
\put(70,77){\line(0,1){20}}
\put(23,88){Fof (\ref{th2:b}) for $\mathcal {H}_{\cap}$:}
\put(34,81){for $\mathcal {O}_{s}$}
\put(76.5,89){Thm. 18}
\multiput(99,88)(-1,0){28}{\line(1,0){0.5}}
\put(72,88){\vector(-1,0){2}}
\put(70,86){\vector(1,0){30}}
\put(71,81.6){Thm. 2~or~18}

\put(100,77){\line(0,1){20}}
\put(100,77){\line(1,0){60}}
\put(100,97){\line(1,0){60}}
\put(160,77){\line(0,1){20}}
\put(103,88){(Nominally robust) UGR:}
\put(120.5,81){$\mathcal {O}_{s}$ for $\mathcal {H}_{\cap}$}
\put(190,87){\vector(-1,0){30}}
\put(168,88){Thm. 3}
\put(161,83){or}

\put(190,77){\line(0,1){20}}
\put(190,77){\line(1,0){60}}
\put(190,97){\line(1,0){60}}
\put(250,77){\line(0,1){20}}
\put(203,88){LaF (\ref{th3:b}) for $\mathcal {H}_{\cap}$:}
\put(213,81){for $\mathcal {O}_{s}$}

\put(33,60){\line(0,1){9}}
\put(229,60){\vector(0,1){9}}
\multiput(227,62)(0,1){7}{\line(0,1){0.5}}
\put(227,62){\vector(0,-1){2}}
\put(175,62){Prop.2.4}
\put(194.3,65){or}
\put(193,60){\vector(0,1){9}}
\put(33,69){\line(1,0){196}}
\put(130,69){\vector(0,1){8}}
\put(131,72){or}
\put(128.1,67.7){$\oplus$}

\put(230,61){Thm. 10~or~11}
\put(209,61){Thm. 11}

\put(0,40){\line(0,1){20}}
\put(0,40){\line(1,0){40}}
\put(0,60){\line(1,0){40}}
\put(40,40){\line(0,1){20}}
\put(8,52){LaSp for $\mathcal {A}$:}
\put(13,44){for $\mathcal {H}_{\cap}$}

\put(75,40){\line(0,1){20}}
\put(75,40){\line(1,0){40}}
\put(75,60){\line(1,0){40}}
\put(115,40){\line(0,1){20}}
\put(78,52){Fof (\ref{th2:b}) for $\mathcal {H}_{\cap}$:}
\put(88,44){for $\mathcal {O}_{b}$}

\put(127,52){Thm. 18}
\multiput(159,51)(-1,0){45}{\line(1,0){0.5}}
\put(117,51){\vector(-1,0){2}}
\put(115,49){\vector(1,0){45}}
\put(122,44.5){Thm. 2~or~18}

\put(160,40){\line(0,1){20}}
\put(160,40){\line(1,0){40}}
\put(160,60){\line(1,0){40}}
\put(200,40){\line(0,1){20}}
\put(173.5,52){UGR:}
\put(169.5,44){$\mathcal {O}_{b}$ for $\mathcal {H}_{\cap}$}

\put(220,40){\line(0,1){20}}
\put(220,40){\line(1,0){40}}
\put(220,60){\line(1,0){40}}
\put(260,40){\line(0,1){20}}
\put(227,52){Stochastic IP}
\put(233,44){for $\mathcal {H}_{\cap}$}

\put(0,0){\line(0,1){20}}
\put(0,0){\line(1,0){150}}
\put(0,20){\line(1,0){150}}
\put(150,0){\line(0,1){20}}
\put(36,11){Matrosov-Foster functions (\ref{1shs:1}) for $\mathcal {H}_{\cap}$:}
\put(70,4){for $\mathcal {O}_{b}$}
\put(120,20){\vector(2,1){40}}
\put(130,21){Thm. 5}
\put(156,34){or}

\put(160,0){\line(0,1){20}}
\put(160,0){\line(1,0){100}}
\put(160,20){\line(1,0){100}}
\put(260,0){\line(0,1){20}}
\put(192.5,11){LaF (\ref{th3:b}) for $\mathcal {H}_{\cap}$:}
\put(203.5,4){for $\mathcal {O}_{b}$}
\put(169,20){\vector(0,1){20}}
\put(170,34){or}
\put(170,21){Thm. 3}

\end{picture}$$
\caption{Summary of stability/recurrence results, where $\mathcal {O}_{b}:=(\mathcal {A}+\varepsilon\mathbb{B}^{\circ})\cup(\mathbb{R}^{n}\setminus(\mathcal {A}+\Delta\mathbb{B}))$ and $\mathcal {O}_{s}:=\mathcal {A}+\varepsilon\mathbb{B}^{\circ}$ for positive real numbers $\varepsilon$ and $\Delta$, ``IP'' denotes ``invariance principle'', ``Laf'' denotes ``Lagrange function'', ``Lyf'' denotes ``Lyapunov function'',  ``Fof'' denotes ``Foster function'', ``LaF'' denotes ``Lagrange-Foster function'', and ``LyF'' denotes ``Lyapunov-Foster function''. See \cite[Fig. 8\&9]{4.1shs} for other summaries of stability/recurrence results.}\label{ttfram}
\end{figure*}

\section{Additional components of stability theory: Sequential compactness, nominal robustness, causality and converse theorems}\label{section4}

\subsection{Sequential compactness}\label{section4.1}


The set of hybrid arcs with closed graphs can be thought of
as a subset in the space of not-identically empty-valued outer semicontinuous set-valued mappings from $\mathbb{R}^2$ to
$\mathbb{R}^n$. As established in \cite[Thm. 5.50, pg. 110\&134\&138\&183]{Rock1998}, equipped with the metric
of graph distance, this space is a separable, locally compact,
complete (and $\sigma$-compact) metric space, which we denote
$(\mathcal {X}, d)$, as presented in \cite{6.1shs} for SDIs, \cite{4.3shs, 3.1shs} for SDDIs, so the case for SDIs obviously holds.


For each $j\in\mathbb{Z}_{\geq0}$, we define $E_{j}:=\mathbb{R}_{\geq0}\times \{0,\ldots, j\}$
and for each hybrid arc $\phi$, we define $\phi|_{E_{j}}$ to be the hybrid arc
with domain $\hbox{dom}\phi\cap E_{j}$ such that $\phi|_{E_{j}}(t, j)=\phi(t, j)$ for
all $(t, j)\in\hbox{dom}\phi|_{E_{j}}$. For each $S\in \mathcal {X}$ we use $S|_{E_{j}}$
to refer to the outer semicontinuous set-valued mapping from $\mathbb{R}^2$ to
$\mathbb{R}^n$ with domain $\hbox{dom} S\cap E_{j}$ such that $S|_{E_{j}}(t, j)=S(t, j)$
for all $(t, j)\in\hbox{dom} S\cap E_{j}$.

\textbf{Assumption 2} \cite{4.3shs}.
The functions $\varphi, \varphi_{i} :\mathcal {X}\rightarrow \mathbb{R}_{\geq0}$, $i\in\mathbb{Z}_{\geq0}$,
are upper semicontinuous and bounded, and for each $\varepsilon>0$
there exist $i^{*}, j^{*}\in\mathbb{Z}_{\geq0}$ such that, for each $i\geq i^{*}$ and each
hybrid arc $\phi$,
\begin{equation}\label{seq1}
\begin{split}
\varphi_{i}(\phi)\leq \varphi_{i}(\phi|_{E_{j}})+\varepsilon, ~\forall~j\geq j^{*}
\end{split}
\end{equation}
and, for each unbounded $\mathcal {N}\subset \mathbb{Z}_{\geq0}$, and each sequence of
locally eventually bounded hybrid arcs $\{S_{i}\}_{i=0}^{\infty}$ and each $j\geq j^{*}$
\begin{equation}\label{seq2}
\begin{split}
\lim_{i\rightarrow\infty,~i\in\mathcal {N}}S_{i}=S\Longrightarrow
\limsup_{i\rightarrow\infty,~i\in\mathcal {N}}\varphi_{i}(S_{i}|_{E_{j}})\leq \varphi(S|_{E_{j}}),
\end{split}
\end{equation}

\textbf{Theorem 12} \cite[Thm. 1]{4.3shs}.
Let Assumptions 1 and 2 hold. Let $\{\hbox{x}_{i}\}_{i=0}^{\infty}$
be an almost surely locally eventually bounded sequence
of solutions, let $\{\Delta_{i}\}_{i=0}^{\infty}$ be a sequence of nonnegative
real numbers, and suppose $\Delta_{i}\leq E[\varphi_{i}(\hbox{x}_{i})]$ for each
$i\in\mathbb{Z}_{\geq0}$. Then there exists a solution $\hbox{x}$ belonging to the
pointwise outer limit of the sequence $\{\hbox{x}_{i}\}_{i=0}^{\infty}$ such that
$\limsup_{i\rightarrow\infty}\Delta_{i}\leq E[\varphi(\hbox{x})]$.

\textbf{Corollary 13} \cite[Coro. 1]{4.3shs}.
Under the conditions of Thm. 12, if the
sequence $\{\hbox{x}_{i}\}_{i=0}^{\infty}$ approaches completeness in probability
then the solution $\hbox{x}$ in the conclusion of Thm. 12 can,
in addition, be chosen to be complete almost surely.

Non-causality means that inside information is leaked, so that exact analysis is based on ``='', compared with the causal case where analysis is based on ``$\in$''. See \cite[Exm. 1]{3.8shs} for that fact that robustness does not holds if the perturbations, although arbitrarily small, are not strictly causal, \cite[Sec. 6.2.3]{3.1shs} and \cite[Exm. 1]{3.3shs} for the fact that the stabilities in probability (UGASp or UGR) may not hold if the causality constraint on solutions is relaxed.

A sequential compactness property for solutions is key to nominal robustness \cite{7.1shs} and thus converse Lyapunov theorems as analyzed in \cite{4.1shs}, \cite{4.3shs} for SDDIs. Also, the results in \cite{4.3shs} for SDDIs as presented above, which use Assumption 1, are crucial for developing the connections between weakly totally recurrent in probability sets and invariant sets, see \cite[Thms. 5\&8, Prop. 2]{6.1shs}, \cite[Thm. 3, Prop. 2]{4.2shs} and \cite[Thm. 2]{4.4shs} for SDDIs. However, such a result for the case where Brownian motion
is involved remains an open problem, as analyzed in \cite[Sec. XI]{5.2shs}, so the nominal robustness and the converse theorems for SDDDIs also remain an open problem.

\subsection{Nominal robustness}\label{section4.2}

Nominal robustness, also called inherent robustness asks how large a disturbance a closed-loop system, designed ignoring uncertainty, can tolerate and still maintain stability. Nominal robustness of stability properties, i.e., stability in probability is
robust to sufficiently small, positive definite state perturbations, is crucial in developing converse Lyapunov theorems as shown in \cite{3.2shs} and \cite{4.6shs}. There exist three kinds of inflations for stochastic hybrid inclusion (\ref{4.5shs:1}), as shown in \cite{7.1shs}, \cite{3.6shs}, \cite{3.1shs}, \cite{5.2shs}, \cite{3.2shs} and \cite{4.6shs}.


Before smoothing the constructed Lyapunov-Foster/Lagrange-Foster/Foster function for (\ref{4.5shs:1}), one initially needs to construct such a function from solutions to an inflation of (\ref{4.5shs:1}).

\emph{\textbf{Case I}: The data of the SHS is modified to slow down recurrence.}

The first inflation is, denoted as $\mathcal {H}_{\nu}$ for SDIs,
\begin{equation}\label{conver1}
x^{+}\in G_{\nu}(x, v):= G(x, v) \cup M_{\nu}(x),
\end{equation}
where $\nu\in \mathcal {K}_{\infty}$, and for compact set $\mathcal {A}$ in \cite{4.6shs},
\begin{equation}\label{conver1.1}
M_{\nu}(x)=\left \{\begin{aligned}
&\emptyset,~~ ~~~~~~~~~~~~~~~x\in \mathcal {A}+2\mathbb{B}, \\
&\mathcal {A}+\nu(|x|_{\mathcal {A}})\mathbb{B},~~\hbox{otherwise},
\end{aligned} \right.
\end{equation}
and for open bounded set $\hat{\mathcal {O}}$ in \cite{3.2shs} satisfying $\hat{\mathcal {O}}+\varepsilon\mathbb{B}\subset \mathcal {O}$,
\begin{equation}\label{conver1.2}
M_{\nu}(x)=\{x_{0}\}+\nu(x-x_{0})\mathbb{B},~~x_{0}\in \hat{\mathcal {O}}.
\end{equation}
The first inflation guarantees radial unboundedness of the
constructed Lagrange-Foster/Foster function.

Similarly, the first inflation of $\mathcal {H}=(C, F,D,G, \lambda, \mu)$ for SDDIs in \cite[Sec. 7]{7.1shs} could be denoted as $\mathcal {H}_{\nu}=(C, F,\mathbb{R}^{n},G_{\nu}, \mu)$ with $M_{\nu}(x)$ satisfying (\ref{conver1.2}) for any $x_{0}\in \mathbb{R}^{n}$, without loss of generality.

The inflation in Case I with (\ref{conver1.1}) is to
possibly slow down but not destroy the convergence to $\mathcal {A}$ of the
solutions that starts far from $\mathcal {A}$, and finally, to result in a radially unbounded
Lyapunov-Foster function in the proof of converse theorem, as shown in \cite{4.6shs} about the converse theorem for UGASp. Similarly, the inflation in Case I with (\ref{conver1.2}) is to preserve recurrence, meanwhile, to make certain quantities grow unbounded in $|x|_{\mathcal {O}}$ since the certification candidate is related
to the expected value of worst case first hitting time to $\mathcal {O}$, as shown in \cite{7.1shs}, which is key to develop converse theorems for recurrence with radially unbounded Foster/Lagrange-Foster functions, see \cite{3.2shs}. Anyway, the inflation in Case I is to prevent a function from not being radially unbounded in developing converse theorem for stabilities. As a result, the nominal robustness is key to converse theorems.

\emph{\textbf{Case II}: Perturbing the system data in a sufficiently small manner.}

The second inflation relative to (\ref{conver1}) in \cite{4.6shs, 3.2shs} is, denoted as $\mathcal {H}_{\rho, \nu}$ for SDIs,
\begin{equation}\label{conver2}
\begin{split}
x^{+}\in G_{\rho, \nu}(x, v):=&\{\omega\in \mathbb{R}^{n}: \omega\in g+\rho(g)\mathbb{B}, \\
& g\in G_{\rho, \nu}(x+\rho(x)\mathbb{B}, v)\},
\end{split}
\end{equation}
where sufficiently small $\rho\in \mathcal {P}\mathcal {D}(\mathcal {A})$ or $\rho\in \mathcal {P}\mathcal {D}(\mathcal {O})$ respectively for compact set $\mathcal {A}$ or open bounded set $\hat{\mathcal {O}}$. Sometimes, $\mathcal {A}$ is defined as the closure of open bounded set $\hat{\mathcal {O}}$.
The second inflation guarantees the smoothness of the constructed Lyapunov function. Such perturbations might include passing from constant to slowly-varying parameters or allowing persistent but small measurement errors in a control system as shown in \cite{3.6shs}. Specifically, the practical stochastic stability properties ensue in the presence of small constant perturbations, i.e., when $\rho\in \mathcal {P}\mathcal {D}(\mathcal {A})$ is a small non-zero constant.

Note that $G(x, v)\subset G_{\nu}(x, v) \subset G_{\rho, \nu}(x, v)$ for all $(x,v)\in \mathbb{R}^{n}\times \mathbb{R}^{m}$ implies that the constructed Lyapunov construction to (\ref{conver2}) is radially unbounded.

The second inflation of $\mathcal {H}$ for SDIs \cite{3.6shs} or SDDIs \cite{3.1shs,5.2shs} not relative to (\ref{conver1}), denoted as $\mathcal {H}_{\delta}$, could be written as
\begin{equation}\label{conver3}
\begin{split}
C_{\delta}:=&\{x\in \mathbb{R}^{n}: (x+\delta(x)\mathbb{B})\cap C \neq \emptyset\},\\
F_{\delta}:=&\overline{\hbox{con}}F(x+\delta(x)\mathbb{B})+\rho(x)\mathbb{B},\\
D_{\delta}:=&\{x\in \mathbb{R}^{n}: (x+\delta(x)\mathbb{B})\cap D \neq \emptyset\},\\
G_{\delta}:=&\{g\in \mathbb{R}^{n}: g=\tilde{g}+\delta(g)\mathbb{B}, \tilde{g}\in G((x+\delta(x)\mathbb{B})\cap D, v)\},
\end{split}
\end{equation}
where $\delta: \mathbb{R}^{n}\rightarrow \mathbb{R}_{\geq0}$ is continuous.

\emph{\textbf{Case III}: Robustness to perturbations of the set.}

In fact, according to \cite{7.1shs} and \cite{3.2shs}, there is another type of inflation, i.e., robust recurrence of an
open bounded set to sufficient small perturbations in the set, i.e., a smaller open set within the original set.

%
%

Note that the three kinds of perturbations are proved to satisfy regularity property, refer to \cite[Fact 3.2]{3.1shs}, \cite[Prop. 2\&3]{4.6shs}, \cite[Prop. 2\&3]{3.2shs}, \cite[Prop. 2]{3.6shs} and \cite[Sec. 7]{7.3shs}. Next we derive the robust stabilities in probability in the following.


Slowing down but not destroying the convergence to $\mathcal {A}$ establishes the robustness of GASp to the inflated data in Case I with (\ref{conver1.1}), as follows borrowed from \cite[Thm. 2]{4.6shs}. See \cite[Thm. 4]{3.2shs} and \cite[Thm. 2]{7.1shs} for the robustness of recurrence to the inflated data in Case I with (\ref{conver1.2}), i.e., making the expected value of worst case first hitting time to $\mathcal {O}$ grow unbounded.

\textbf{Theorem 14} (Case I: Slowing down the convergence) \cite[Thm. 2]{4.6shs}.
If the compact set $\mathcal {A}\subset \mathbb{R}^{n}$ is GASp for SDIs/SDDIs in (\ref{4.5shs:1}) then
there exists $\nu\in \mathcal {K}_{\infty}$ such that $\mathcal {A}$ is GASp for (\ref{conver1}) and (\ref{conver1.1}).

Perturbing the system data in a sufficiently small manner establishes the robustness of GASp to measurement noise, additive disturbances and parametric uncertainty in the system data, as follows borrowed from \cite[Thm. 3]{4.6shs}. Refer to \cite[Thm. 5]{3.2shs} and \cite[Thm. 3]{7.1shs} for the similar results of robust recurrence.

\textbf{Theorem 15} (Case II: Perturbations of system data) \cite[Thm. 3]{4.6shs}.
If the compact set $\mathcal {A}\subset \mathbb{R}^{n}$ is GASp for SDIs/SDDIs in (\ref{4.5shs:1}) then
there exists $\rho\in \mathcal {P}\mathcal {D}(\mathcal {A})$ such that $\mathcal {A}$ is GASp for (\ref{conver2}).


we now show the recurrence of an open bounded set implies recurrence of a smaller open set within the original set, i.e., robustness to perturbations in the set, borrowed from \cite[Thm. 1]{7.1shs}. See \cite[Thm. 3]{3.2shs} for similar results of robust recurrence.

\textbf{Theorem 16} (Case III: Perturbations in the set) \cite[Thm. 1]{7.1shs}.
Let the open bounded set $\mathcal {O}\subset \mathbb{R}^{n}$ be globally recurrent for SDIs/SDDIs in (\ref{4.5shs:1}).
Then, there exists an open bounded set $\hat{\mathcal {O}}$ and
$\varepsilon>0$ such that $\hat{\mathcal {O}}+\varepsilon\mathbb{B}\subset \mathcal {O}$ and $\hat{\mathcal {O}}$ is globally recurrent for SDIs/SDDIs in (\ref{4.5shs:1}).


\subsection{Converse theorems}\label{section4.3}

Converse theorem means Lyapunov-based necessary and sufficient conditions for asymptotic stability in probability or recurrence in \cite{4.6shs}, \cite{3.2shs}.

\textbf{Theorem 17} \cite[Thm. 1]{4.6shs}.
If the compact set $\mathcal {A}\subset \mathbb{R}^{n}$ is GASp for $\mathcal {H}= (C\cup D,G, \mu)$ then
there exists a smooth on $\mathbb{R}^{n}\setminus \mathcal {A}$, continuous, (regular) Lyapunov-Foster function relative to $\mathcal {A}$ for (\ref{4.5shs:1}).


Thm. 17 is a converse of \cite[Thm. 1\&2]{3.3shs} or \cite[Thm. 4.5]{3.1shs} with the same hybrid model, establishing that Lyapunov-based conditions (Lyapunov-Foster function in (\ref{th4:b})) imply GASp for (\ref{4.5shs:1}) as shown in Thm. 1 and that a continuous Lyapunov-Foster function is a sensible
stability analysis tool for SHIs. Also, it shows in \cite[Thm. 1\&3]{3.6shs}, \cite[Th. 1\&2]{3.8shs} and \cite[Prop. 1]{3.2shs} that how to pass from a continuous Lyapunov/Foster function $V$ for $\mathcal {H}= (D,G, \mu)$ to a
continuous Lyapunov/Foster function $\Gamma(V)$, where $\Gamma\in \mathcal {K}_{\infty}$ is an appropriate
concave function, for the sufficiently small, state-dependent, strictly causal, worst-case
perturbations. In fact, an appropriate concave function $\Gamma\in \mathcal {K}_{\infty}$ is necessary
to avoid the integral of function $V$ being unbounded, as shown in \cite[Exm. 3]{3.6shs}.

Next, we present the converse theorem for UGR, as shown in \cite{3.2shs}.

\textbf{Theorem 18} \cite[Thm. 2]{3.2shs}.
The open bounded set $\mathcal {O}\subset \mathbb{R}^{n}$ is strongly globally
recurrent for $\mathcal {H}= (D,G, \mu)$ if and only if there exists a smooth Foster function relative to $\mathcal {O}$ for $\mathcal {H}= (C\cup D,G, \mu)$.

Note that here we did not distinguish the sufficient recurrence-Lyapunov function (i.e., Foster function) and recurrence-Lyapunov function as analyzed in \cite{3.2shs}, since the latter can be regarded as an (tighter/stronger) extension of Foster function, in order to uniformly
bound the expected value of the function for solutions starting from $\mathcal {O}$. Moreover, \cite{3.2shs} shows that recurrence-Lyapunov function implies sufficient recurrence-Lyapunov function, but not vise versa \cite[Exm. 1]{3.2shs}; similarly, Lagrange-Foster function implies Foster function \cite[Thm. 4.4]{3.1shs}, but not vise versa. In the above theorems, nominal robustness is used to smooth the established function in the final step, refer to \cite[Thm. 1]{4.6shs} and \cite[Th. 2]{3.2shs}.

\section{applications}\label{section5}


Stochastic hybrid inclusion (\ref{4.5shs:1}), including linear time-triggered SHSs and state-driven SHSs, can be used to model sampled-data stochastic control systems, average dwell-time switching among stochastic differential equations, model spontaneous transitions and a stochastic bouncing ball system as shown in \cite{5.2shs}, \cite{3.1shs}. On the other hand, the stochastic hybrid approach can be used to achieve robustness, i.e., designing hybrid control law, for example, the robust synchronization on the circle in the presence of adversaries while maintaining good convergence rates in \cite{3.4shs}, \cite{3.7shs}, which is extended to robust global attitude control in \cite{2011tac}. Those reasons inspire researchers to delve more deeply into SHSs, including the models and potentially significant advantages for further applications.


Next, we present two typical examples for a stochastic bouncing ball system and the robust synchronization on the circle to show the advantages and necessity of SHIs.

\subsection{Stochastic bouncing ball}

First, we present the stability in probability using the relax sufficient conditions in Sec. \ref{section3.3}, i.e., a family of Matrosov-Foster functions \cite[Sec. 6.1]{3.1shs}, \cite[XII-C]{5.2shs} and stochastic invariance principle \cite[Exm. 4]{4.2shs}, respectively.

The bouncing ball is a system of the form $\mathcal {H}=(C,F,D,G,\mu)$ for SDDIs (\ref{4.5shs:1}) with data
\begin{equation}\label{bbc1}
\begin{split}
C :=&\{x\in \mathbb{R}^{2}: x_{1}\geq0\},\\
F(x):=&(x_{2}, -\gamma)^{T},~\forall ~x\in C\setminus \{(0,0)\},\\
F(0):=& \{0\}\times [-\gamma, 0],\\
D:=&\{x\in \mathbb{R}^{2}: x_{1}=0,~x_{2}\leq0\} ,\\
G_{g}(x,v):=&-\sigma(v)x+(0, [0,\gamma])^{T},\\
\end{split}
\end{equation}
where the $\sigma: \mathbb{R}\rightarrow\mathbb{R}_{\geq0}$ is measurable and $\eta\mathbb{R}_{\geq0}$. Note that the interval $[0,\eta]$ is added to $-\sigma(v)x_{2}$ in the $x^{+}_{2}$ inclusion, while $x^{+}_{1}=0$ for $x \in D$. The data $(C,F,D,G_{g})$ satisfies Assumption 1 for each $\eta\mathbb{R}_{\geq0}$. The case $\eta=0$ we call the bouncing ball without offset. We assume there exists $r>0$ such that
\begin{equation}\label{bbc2}
\begin{split}
\lambda:=\int_{\mathbb{R}}\sigma^{r}(v)\mu(dv)<1,
\end{split}
\end{equation}
where $r$ can be chosen as 1 or 2 as analyzed in \cite[Sec. 6.1]{3.1shs}.

Next we analyze the stabilities in probability: UGASp without offset and UGR with offset.
Compared with the deterministic bouncing ball in \cite[pg. 2\&35\&54\&61\&82\&125\&178]{2012teel} using deterministic invariance principle \cite[pg. 174]{2012teel}, here we could also apply the Matrosov-Foster functions, besides the stochastic invariance principle \cite[Exm. 4]{4.2shs}, to verify UGASp for (\ref{bbc1})-(\ref{bbc2}).

\subsubsection{UGASp of origin without offset:}

Let $\mathcal {A}$ be the origin in $\mathbb{R}^{2}$ and consider the certification
candidate $V(x)=(x^{2}_{2}+2\gamma x_{1})^{\frac{r}{2}}$ for all $x\in \mathbb{R}^{2}$. It is easy to check that $V(x)$ satisfy (\ref{th1:A}) and (\ref{th1:B}) with $\varpi(x)=|x|_{\mathcal {A}}$ for some functions $\ln(V(x)+1)=:\alpha_{1}\in \mathcal {K}_{\infty}$ and $\exp(V(x))-1=:\alpha_{2}\in \mathcal {K}_{\infty}$.

For all $x\in C$,
\begin{equation}\label{b3}
\begin{split}
\langle V(x), f(x)\rangle=&\frac{r}{2}(x^{2}_{1}+2\gamma x_{1})^{-\frac{r}{2}}(2x_{2}\dot{x}_{2}+2\gamma \dot{x}_{1})\\
=&\frac{r}{2}(x^{2}_{1}+2\gamma x_{1})^{-\frac{r}{2}}(2x_{2}(-\gamma)+2\gamma x_{2})=0,
\end{split}
\end{equation}
while for $x\in D$
\begin{equation}\label{b4}
\begin{split}
&\int_{\mathbb{R}}\sup_{g\in G_{0}(x,v)}V(g)\mu(dv)\\
=&\int_{\mathbb{R}}\sup_{g\in G_{0}(x,v)}((-\sigma(v)x_{2})^{2}+0)^{\frac{r}{2}}\mu(dv)\\
=&|x_{2}|^{r}\int_{\mathbb{R}}\sigma^{r}(v)\mu(dv)\\
=&\lambda |x_{2}|^{r}=\lambda ((x^{2}_{2}+2\gamma *0)^{\frac{r}{2}})=\lambda V(x).
\end{split}
\end{equation}
So $\langle V(x), f(x)\rangle=0$ for
all $x\in C$ satisfying (\ref{th1:C}), and $\int_{\mathbb{R}}\sup_{g\in G_{0}(x,v)}V(g)\mu(dv) \leq \lambda V(x)\leq V(x)$ for $x\in D$ imply that $V$ is a Lyapunov function for the origin. According to Thm. 4, the origin is UGSp for $(C,F,D,G_{0},\mu)$.

\emph{1.1) Using Matrosov-Foster functions \cite[Sec. 6.1]{3.1shs}, \cite[XII-C]{5.2shs}.}

Let $K\subset \mathcal {A}^{c}$ be compact. Let candidate functions be $V0(x)=V(x)$ and $V1(x) = \beta_{1}+x_{2}$
where $\beta_{1}+x_{2}\geq 0$ for all $x_{2}\in K$. Let $\beta_{2}> 0$ be such that $\beta_{1}+x_{2}\leq \beta_{2}$
for all $x_{2}\in K$. Based on (\ref{b3})-(\ref{b4}), we have
\begin{equation}\label{b5}
\begin{split}
&\langle V_{0}(x), f(x)\rangle=0,\\
&\int_{\mathbb{R}}\sup_{g\in G_{0}(x,v)}V_{0}(g)\mu(dv) \\
&\leq \lambda V_{0}(x)=V_{0}(x)+(\lambda-1)V_{0}(x)=V_{0}(x)+u_{d,0},
\end{split}
\end{equation}
where $u_{c,0}=0$ and $u_{d,0}=(\lambda-1)V_{0}(x)<0$.
Similarly, choosing $r=1$, we get
\begin{equation}\label{b5}
\begin{split}
&\langle V_{1}(x), f(x)\rangle=\dot{x}_{2}=-\gamma<0,\\
&\int_{\mathbb{R}}\sup_{g\in G_{0}(x,v)}V_{1}(g)\mu(dv)\\
&=\int_{\mathbb{R}}\sup_{g\in G_{0}(x,v)}(\beta_{1}+(-\sigma(v)x_{2}))\mu(dv)\\
&=\int_{\mathbb{R}}\sup_{g\in G_{0}(x,v)}(\beta_{1}+\sigma(v)|x_{2}|)\mu(dv)\\
&\leq\beta_{1}+|x_{2}|\int_{\mathbb{R}}\sup_{g\in G_{0}(x,v)}\sigma(v)\mu(dv)\\
&\leq\beta_{1}+|x_{2}|\lambda\\
&=\beta_{1}+x_{2}+|x_{2}|-x_{2}\\
&=V_{1}(x)+|x_{2}|-x_{2}\\
&\leq V_{1}(x)+|x_{2}|+\beta_{1}\leq V_{1}(x)+\beta_{2}=V_{1}(x)+u_{d,2},
\end{split}
\end{equation}
where $u_{c,1}=-\gamma<0$ and $u_{d,0}=\beta_{2}>0$.
As a result, $(V_{0},_{V}1)$ are Foster-Matrosov functions for $\mathbb{R}^{2}\setminus K$, so Prop. 1 implies that $\mathcal {A}+\varepsilon \mathbb{B}^{\circ}$ is UGR since $\mathcal {K}_{\infty}\subset\mathcal {G}_{\infty}$. According to
Coro. 7, the origin is UGASp for $(C,F,D,G_{0},\mu)$ in (\ref{bbc1})-(\ref{bbc2}).

\emph{1.2) Using invariance principle \cite[Exm. 4]{4.2shs}.}

Instead of Foster-Matrosov functions, we use Thm. 8 to establish UGASp of the origin,
i.e., ruling out the existence of a
compact subset of $K_{\delta, \Delta}$
that is weakly forward invariant
in probability. The analysis in \cite[pg. 35]{2012teel} implies that for the bouncing ball,
every complete solution is Zeno, i.e., having an infinite number of jumps. The above analysis for $\langle V(x), f(x)\rangle=0$ in (\ref{b3}) shows that the Lyapunov function $V(x)$ remains constant during flows and
(\ref{b4}) implies that
$\int_{\mathbb{R}}\sup_{g\in G_{0}(x,v)}V(g)\mu(dv)=\lambda V(x)$ for all $x\in D$, where $\lambda<1$.
Therefore, classical results for discrete-time stochastic systems give that $V$ almost surely
converges to zero. So, the set $K_{\delta, \Delta}$ for each $0<\delta<\Delta<\infty$ contains no compact set that is almost
surely weakly forward invariant for $(C,F,D,G_{0},\mu)$ in (\ref{bbc1})-(\ref{bbc2}). Consequently, Thm. 8 implies that the origin is UGASp for $(C,F,D,G_{0},\mu)$ in (\ref{bbc1})-(\ref{bbc2}).

Next we present the UGR of origin with offset \cite[Sec. 6.1]{3.1shs}.

\subsubsection{UGR of origin with offset:}

According to above analysis, the same function $V$ is a Lagrange function for $\bar{\mathcal {O}}$ where $\mathcal {O}:=\tilde{\gamma}\eta \mathbb{B}^{\circ}$ for $(C,F,D_{\cap},G_{\eta},\mu)$, so $\bar{\mathcal {O}}$ is LaSp since $\mathcal {K}_{\infty}\subset\mathcal {G}_{\infty}$.
Based on the same Foster-Matrosov functions $(V_{0},_{V}1)$ for $\mathbb{R}^{2}\setminus K$, Prop. 1 implies that $\mathcal {O}$ is UGR for $(C,F,D,G_{g},\mu)$ in (\ref{bbc1})-(\ref{bbc2}).


\subsection{Robust global almost sure synchronization on the circle}
Nominal global almost sure synchronization result is established in \cite{3.7shs}, where a robustness property is established with respect to
sufficiently small, worst-case perturbations, based on the stochastic, hybrid state feedback
algorithm. In contrast, non-stochastic almost globally asymptotically synchronizing algorithms do not confer this robustness property.

Continuous stabilizers for almost global synchronization
and discontinuous stabilizers for global synchronization on
the circle exhibit non-robust stability
behavior in the presence of adversaries. Stochastic gossip algorithms to achieve almost
sure asymptotic stability tend to have slow convergence rates.
This motivates the stochastic hybrid approach in \cite{3.4shs} and \cite{3.7shs} to
achieve robust synchronization on the circle in the presence
of adversaries while maintaining good convergence rates.
In \cite{3.7shs} a stochastic hybrid algorithm for synchronizing
two agents on the circle is proposed, while an extension of the
algorithm to a finite number of agents under all-to-all communication is discussed and proved in \cite{3.4shs}
by providing a complete Lyapunov function, where robustness margins are also established.



\section{Conclusion}\label{section6}

For SHIs, besides the possibly stochastic continuous evolution via a stochastic differential inclusions, randomness is mainly restricted to the jump map based on forced transitions while sometimes also to the spontaneous transitions in the difference inclusions.
Mild regularity conditions on the system data are summarized to guarantee existence of random solutions, the equivalence between non-uniform and uniform versions of stochastic stability properties, and the nominal robustness of Lyapunov conditions.
We have presented several properties for SHIs, including random solutions, stabilities and recurrence as well as corresponding (relaxed) Lyapunov-based sufficient/necessary conditions, sequential compactness, nominal robustness and converse theorems.
Besides the sufficient Lyapunov conditions, other weaker conditions are summarized, as stochastic invariance principle and Matrosov-Fosters functions. The Lyapunov conditions are similar to classical Lyapunov conditions, rather than based on the derivative of the expected value of a function. An explicit example for stochastic bouncing ball system is present to show the stability results.

Besides the above mentioned applications in Section \ref{section5}, SHIs can be potentially used to
1) design hybrid control law for networked rigid-body attitude motions \cite{2022zddtac}, where one needs to simultaneously consider the hybrid control law and the discontinuous communications;
2) model and analyze a class of nonlinear systems with stochastic random reset controller \cite{2023zddauto};
3) model and analyze the networked security problems using event-triggering sampling control strategy under the digital communication channel, including but not limited to multi-agent systems \cite{2023sandratac2};
4) model and analyze for a class of unknown stochastic hybrid inclusions with Gaussian process regression \cite{2019Sandra-auto}, \cite{2019sandra-tac}, \cite{2013sandra-cdc};
5) model and analyze other systems, including financial systems, air traffic management systems, biological systems and power systems, etc, as analyzed in \cite{4.1shs}.

\section*{Acknowledgment}
The authors would like to thank Prof. Andrew R. Teel for his guidance on stochastic hybrid systems.











\end{document}